\documentclass[aps,prd,nofootinbib,twocolumn,superscriptaddress]{revtex4-2}

\usepackage{amsmath,amssymb,amsthm,amstext,mathrsfs}
\usepackage{natbib}
\usepackage{graphicx}
\usepackage{color}
\usepackage[dvipsnames]{xcolor}
\usepackage{array,enumerate}
\usepackage{bm}
\usepackage{multirow}
\usepackage[breaklinks,colorlinks,citecolor=cyan,urlcolor=NavyBlue]{hyperref}
\usepackage{braket}
\usepackage{txfonts}
\usepackage{physics}
\usepackage{dcolumn}

\usepackage{tikz}
\usetikzlibrary{decorations.pathmorphing}
\tikzset{snake it/.style={decorate, decoration=snake}}

\def\be{\begin{equation}}
\def\ee{\end{equation}}
\def\ba{\begin{aligned}}
\def\ea{\end{aligned}}
\def\bpm{\begin{pmatrix}}
\def\epm{\end{pmatrix}}

\newcommand{\dif}{\text{d}}
\newcommand{\JJ}{\mathcal{J}}
\newcommand{\FF}{\mathcal{F}}

\begin{document}

\title{Axion corrections to photon superradiant scattering by Kerr black holes}

\author{Zhi-Qing Zhu}
\email{zhuzhiqing24@mails.ucas.ac.cn}
\affiliation{International Centre for Theoretical Physics Asia-Pacific,
University of Chinese Academy of Sciences, 100190 Beijing, China}

\author{Jun Zhang}
\email{zhangjun@ucas.ac.cn}
\affiliation{International Centre for Theoretical Physics Asia-Pacific,
University of Chinese Academy of Sciences, 100190 Beijing, China}
\affiliation{Taiji Laboratory for Gravitational Wave Universe,
University of Chinese Academy of Sciences, 100049 Beijing, China}

\begin{abstract}
We study photon superradiant scattering by a Kerr black hole in the presence of an axion-photon coupling and investigate how this coupling modifies the electromagnetic amplification factor. We solve the coupled axion and electromagnetic equations in a fully relativistic framework. We distinguish two regimes according to the relation between the sourced axion frequency and the axion mass. In the radiative regime, the axion can propagate to spatial infinity and carry away energy, thereby suppressing photon amplification. In the subthreshold regime, the axion is confined by the mass barrier and exchanges energy with the black-hole horizon through quasibound-state excitations. In this regime, photon amplification can be either enhanced or suppressed, depending on whether the relevant axion modes satisfy the superradiance condition.
\end{abstract}
\maketitle

\section{Introduction}

Black hole superradiance is one of the cleanest manifestations of energy extraction from a rotating dissipative system.  For a bosonic mode with time and azimuthal dependence $e^{-i\omega t+im\varphi}$, the flux through the event horizon of a Kerr black hole becomes negative when $0<\omega<m\Omega_H$, so that the reflected wave carries more energy than the incident wave.  The mechanism was first identified for rotating absorbers and was soon developed in the black-hole context~\cite{ZelDovich:1971,Starobinskii:1973vzb,Starobinskil:1974nkd,Press:1972zz}. The separability of the Kerr perturbation equations then made it possible to calculate amplification factors for scalar, electromagnetic, and gravitational waves in a fully relativistic setting~\cite{Teukolsky:1973ha,Teukolsky:1974yv}. Superradiant scattering and its extensions have since become standard probes of horizon dissipation and rotating-spacetime dynamics~\cite{BekensteinSchiffer:1998,Brito:2015oca}; electromagnetic superradiance has also been examined for waves generated by astrophysical sources near the black hole~\cite{KobayashiTomimatsu:2010}.

For a massive bosonic field, the mass term supplies a natural confining potential and turns repeated superradiant amplification into an instability. The resulting quasibound states (QBSs) have hydrogenic structure in the weak-binding regime, while their small imaginary frequencies encode either growth by superradiant extraction or decay by absorption at the horizon \cite{Zouros:1979iw,Detweiler:1980uk,Dolan:2007}.  This observation has attracted renewed attention because ultralight bosons, such as axions and axion-like particles, can form macroscopic ``gravitational atoms'' around astrophysical black holes, thereby converting measurements of black-hole spins, gravitational waves, and binary dynamics into probes of particle physics \cite{Arvanitaki:2009fg,Arvanitaki:2010sy,Arvanitaki:2014wva,Brito:2014wla,Baumann:2018vus,Zhang:2018kib,Zhang:2019eid}. 

The effects of interactions on superradiance of a massive boson field have been discussed in the literature~\cite{Zhu:2025enp,Spieksma:2023vwl,Chen:2019fsq,Chen:2021lvo,Chen:2022oad,Chen:2023vkq,Lyu:2025lue,Xie:2025npy,Yang:2023vwm,Baryakhtar:2020gao}. Here we address the complementary scattering problem. We take an electromagnetic wave incident from spatial infinity as the primary superradiant field, assume no background axion configuration, and ask how the axion field dynamically sourced during the scattering process feeds back on the outgoing photon. This distinction is important because the relevant axion response depends qualitatively on whether its frequency lies above or below the mass threshold. Above the threshold, the axion can carry energy to spatial infinity through outgoing radiation. Below the threshold, it is evanescent at infinity and its response is controlled by QBSs and by energy exchange with the horizon. Capturing both regimes consistently requires the Kerr scattering problem, its radiative boundary conditions, and the effectively non-Hermitian nature of black hole rather than only a Newtonian bound-state description. Useful ingredients for this treatment include frequency-domain Kerr Green functions and recently developed conserved bilinear forms and relativistic perturbation theory for black hole modes \cite{Leaver:1985ax,leaver1986solutions,Yang:2013shb,Green:2022htq,Cannizzaro:2023jle,Cannizzaro:2025vpb,Fu:2025ztk}.

This paper is organized as follows. In Sec.~\ref{sec:bosonicfield}, we review the spin-$1$ Teukolsky equation for the electromagnetic field and the massive Klein--Gordon equation for the axion on Kerr spacetime. In Sec.~\ref{sec:superradiance}, we construct the perturbative axion--photon system and compute the correction to photon superradiance in the radiative and sub-threshold regimes. We summarize our results in Sec.~\ref{sec:conclusion}. In this paper, we use units with $G=c=\hbar=1$.

\section{Bosonic Fields in Kerr Spacetime}
\label{sec:bosonicfield}

In this section, we review the wave equations governing bosonic fields on a fixed Kerr background. In Boyer-Lindquist coordinates, the metric of a Kerr black hole with mass $M$ and spin $a\equiv J/M$ is given by
\be
\ba
\dif s^2
=&-\left(1-\frac{2Mr}{\Sigma}\right)\dif t^2
-\frac{4Mar\sin^2\theta}{\Sigma}\dif t\dif \varphi \\
&+\frac{\Sigma}{\Delta}\dif r^2+\Sigma \dif \theta^2
+\frac{(r^2+a^2)^2-\Delta a^2\sin^2\theta}{\Sigma}
\sin^2\theta\,\dif \varphi^2,
\ea
\ee
where $\Sigma = r^2+a^2\cos^2\theta$ and $\Delta = r^2-2Mr+a^2$. The event horizon is located at $r_+ = M+\sqrt{M^2-a^2}$, and the angular velocity of the horizon is $\Omega_H = a/2Mr_+$. We focus on two cases relevant to our analysis: a massless spin-1 field, corresponding to the electromagnetic field and governed by the spin-1 Teukolsky equation, and a massive scalar field, corresponding to the axion field and governed by the massive Klein-Gordon equation on a Kerr background.

\subsection{Massless spin-1 field}

To derive the spin-1 Teukolsky equation, we first introduce the Kinnersley tetrad in Boyer-Lindquist coordinates,
\be
\ba
l^\mu &= \left(r^2+a^2/\Delta,1,0,a/\Delta\right),\\
n^\mu &= \left(r^2+a^2,-\Delta,0,a\right)/2\Sigma,\\
m^\mu &= \left(i a\sin\theta,0,1,i/\sin\theta\right)/
2^{1/2}(r+i a\cos\theta), 
\ea
\ee
together with $m^\mu$'s complex conjugate $\bar m^\mu$, which is normalized as $l^\mu n_\mu=-1$ and $m^\mu \bar m_\mu=1$. Then the complex Newman-Penrose (NP) scalars can be obtained by projecting the real electromagnetic field strength $F_{\mu\nu}$ onto the tetrad,
\be
\ba
\phi_0 = F_{\mu\nu}l^\mu m^\nu,
\phi_1 = \frac{1}{2}F_{\mu\nu} \left(l^\mu n^\nu+\bar m^\mu m^\nu\right),
\phi_2 = F_{\mu\nu}\bar m^\mu n^\nu .
\ea
\ee
Conversely, the electromagnetic field strength can be reconstructed from the NP scalars,
\be
    F^{\mu\nu} = 2\left[\phi_1 (n^{[\mu} l^{\nu]} + m^{[\mu} \bar{m}^{\nu]}) + \phi_2 l^{[\mu} m^{\nu]} + \phi_0 \bar{m}^{[\mu}n^{\nu]} \right] + \mathrm{c.c.},
\ee
and the dual tensor is given by
\be\label{eq:dualF}
    \tilde F_{\mu\nu} = 2i \left[\phi_1 (n_{[\mu} l_{\nu]} + m_{[\mu} \bar{m}_{\nu]}) + \phi_2 l_{[\mu} m_{\nu]} + \phi_0 \bar{m}_{[\mu}n_{\nu]} \right] + \mathrm{c.c.},
\ee
where square brackets denote antisymmetrization on subscripts. With these definitions, the two electromagnetic Lorentz invariants can be expressed in terms of the NP scalars as
\be
\ba\label{eq:FF}
    &F_{\mu\nu}F^{\mu\nu} = 8\, \mathrm{Re}\,(\phi_0 \phi_2 - \phi_1^2),\\
    &F_{\mu\nu}\tilde F^{\mu\nu} = 8\, \mathrm{Im}\,(\phi_0 \phi_2 - \phi_1^2).
\ea
\ee

Using Maxwell's equations $\nabla_\mu F^{\mu\nu}=J^\nu, \nabla_{[\lambda}F_{\mu\nu]}=0$, one can eliminate $\phi_1$ and obtain two decoupled equations for $\phi_0$ and $\phi_2$. By further defining
\be
\psi_{+1}=\phi_0,\qquad
\psi_{-1}=(r-i a\cos\theta)^2\phi_2,
\ee
the equations of $\phi_0$ and $\phi_2$ can be written in the same form, i.e., the Teukolsky equation with spin weight $s=\pm 1$:
\be\label{eq:Teukolsky}
\mathcal{H}_{s}\psi_{s}=4\pi\Sigma T_{s},
\ee
where

\be
\ba\label{eq:TeukolskyH}
\mathcal{H}_{s} =& \left[
\frac{(r^2+a^2)^2}{\Delta}-a^2\sin^2\theta
\right]\partial_t^2
+\frac{4Mar}{\Delta}\partial_t\partial_\varphi
\\
&
+\left(
\frac{a^2}{\Delta}-\frac{1}{\sin^2\theta}
\right)\partial_\varphi^2
-\Delta^{-s}\partial_r
\left(\Delta^{s+1}\partial_r\right)
\\
&
-\frac{1}{\sin\theta}\partial_\theta
\left(\sin\theta\,\partial_\theta\right)
-2s\left[
\frac{a(r-M)}{\Delta}
+\frac{i\cos\theta}{\sin^2\theta}
\right]\partial_\varphi
\\
&
-2s\left[
\frac{M(r^2-a^2)}{\Delta}-r-i a\cos\theta
\right]\partial_t
+\left(s^2\cot^2\theta-s\right) \,,
\ea
\ee
and $T_s$ denotes the potential sources. For $T_s=0$, we take the ansatz 
\be\label{eq:EMansatz}
\psi_s= \mathcal{A}_\gamma e^{-i\omega t+im\varphi}S_{s\ell m}(\theta)R_{s\ell m}(r),
\ee
where $\mathcal{A}_\gamma$ is an overall amplitude of the electromagnetic field. Eq.~\eqref{eq:Teukolsky} therefore can be separated into angular and radial equations
\be
\ba\label{eq:angular}
&\frac{1}{\sin\theta}\frac{\dif}{\dif\theta}
\left(\sin\theta\frac{\dif S_{s\ell m}}{\dif\theta}\right)
+\Bigg[a^2\omega^2\cos^2\theta-2a\omega s\cos\theta
\\
&\qquad
-\frac{(m+s\cos\theta)^2}{\sin^2\theta}
+s+A_{s\ell m}\Bigg]\, S_{s\ell m}=0.
\ea
\ee
and
\be
\ba\label{eq:radial}
&\Delta^{-s}\frac{\dif}{\dif r}
\left(\Delta^{s+1}\frac{\dif R_{s\ell m}}{\dif r}\right)
+\Bigg[\frac{K^2-2is(r-M)K}{\Delta}
\\
&\qquad
+4is\omega r-\lambda_{s\ell m}\Bigg]\, R_{s\ell m}=0,
\ea
\ee
where $\omega$ is the photon frequency and $m$ is the azimuthal number with $K=(r^2+a^2)\omega-am$ and $\lambda_{s\ell m}=A_{s\ell m}+a^2\omega^2-2am\omega$.

Subject to the regularity boundary conditions, the solutions of the angular equation~\eqref{eq:angular} are spin-weighted spheroidal harmonic satisfying the orthogonality condition
\be\label{eq:angularnorm}
\int_0^\pi \dif\theta\,\sin\theta\,
S_{s\ell m}(\theta;\, a\omega)S^*_{s\ell' m}(\theta; \, a \omega)= \frac{1}{2\pi} \delta_{\ell\ell'}.
\ee
For the equation~\eqref{eq:radial}, it is conventional to introduce the in-mode and the up-mode as the two linearly independent homogeneous solutions. Specifically, the in-mode is purely ingoing at the black hole horizon, and its asymptotic behavior is given by
\be\label{eq:innorm}
\ba
R^{\rm in}_{s\ell m} \to
\begin{cases}
    \Delta^{-s}e^{-i k_H r_*}, \qquad &r\to r_+ \\
    A^{\rm in}_{s\ell m}r^{-1}e^{-i\omega r_*} + A^{\rm out}_{s\ell m}r^{-2s-1}e^{i\omega r_*}, \qquad &r\to\infty
\end{cases} \,,
\ea
\ee
where $r_*$ is the tortoise coordinate defined by $dr_*/dr=(r^2+a^2)/\Delta$, $k_H=\omega-m\Omega_H$, and $A^{\rm in/out}_{s\ell m}$ are the incident and reflection amplitudes at infinity. The transmitted amplitude at horizon is set to unity as a normalization convention. The up mode, on the other hand, is purely outgoing at spatial infinity.  We
normalize it similarly by setting the outgoing amplitude at infinity to unity, so that its asymptotic behavior is given by
\be\label{eq:upnorm}
\ba
R^{\rm up}_{s\ell m} \to
\begin{cases}
    A^{\rm down}_{s \ell m} \Delta^{-s}e^{-i k_H r_*} + A^{\rm up}_{s \ell m} e^{i k_H r_*}, \qquad &r\to r_+\\
    r^{-2s-1}e^{i\omega r_*}, \qquad &r\to\infty 
\end{cases} .
\ea
\ee

The radiative degrees of freedom of the electromagnetic field can be encoded in either of the two decoupled NP scalars, $\phi_0$ or $\phi_2$. The equivalence between the two descriptions is made explicit by the Teukolsky-Starobinsky identities \cite{Teukolsky:1974yv,Starobinskii:1973vzb,Starobinskil:1974nkd},
\be
\ba
    &\mathscr{D}\mathscr{D}R_{-1}=\frac{1}{2}R_{+1},\quad
    \Delta\mathscr{D}^\dagger\mathscr{D}^\dagger\Delta R_{+1}=2B^2R_{-1}, \\
    &\mathscr{L}_0\mathscr{L}_1S_{+1}=BS_{-1},\quad
    \mathscr{L}_0^\dagger\mathscr{L}_1^\dagger S_{-1}=BS_{+1},
\ea
\ee
where
\be
\ba
\mathscr{D} &= \partial_r - i\frac{K}{\Delta},\qquad
\mathscr{D}^\dagger = \partial_r + i\frac{K}{\Delta},\\
\mathscr{L}_n &= \partial_\theta + m\csc\theta - a\omega\sin\theta + n\cot\theta, \\
\mathscr{L}_n^\dagger &= \partial_\theta - m\csc\theta + a\omega\sin\theta + n\cot\theta, \\
B &= \left[ \lambda_{s \ell m}+s(s+1) \right]^2+4am\omega-4a^2\omega^2.
\ea
\ee

In order to fix the relative normalization between the spin-$\pm 1$ solutions, we introduce the normalized NP scalars $\hat{\phi}_i$ as
\be\label{eq:phi0phi2}
\ba
\hat{\phi}_0 &= e^{-i\omega t+im\varphi} S_{+1}(\theta) R_{+1}(r),\\
\hat{\phi}_2 &= \frac{B}{(r-i a\cos\theta)^2} e^{-i\omega t+im\varphi} S_{-1}(\theta) R_{-1}(r),
\ea
\ee
such that $S_{\pm 1}$ and $R_{\pm 1}$ satisfy the normalization conditions given in Eqs.~\eqref{eq:angularnorm}, \eqref{eq:innorm} and \eqref{eq:upnorm}. With the relative normalization fixed as above, the spin-$+1$ and spin-$-1$ asymptotic amplitudes are related by
\be
    A_{1\ell m}^{\rm in} = -\frac{8\omega^2}{B} A_{-1\ell m}^{\rm in}, \quad A_{1\ell m}^{\rm out} = -\frac{B}{2\omega^2} A_{-1\ell m}^{\rm out}
\ee

The remaining NP scalar $\phi_1$ is not an independent degree of freedom. Once the spin-$\pm1$ radial and angular functions are specified, $\phi_1$ can be derived from Maxwell's equations. With the same normalization convention as above, it can be written as
\be\label{eq:phi1}
\ba
\hat{\phi}_1 &= -\frac{e^{-i\omega t+im\varphi}}{2(r-i a\cos\theta)^2} \left[g_{+1}\mathscr{L}_1S_{+1}+f_{-1}\mathscr{D}R_{-1}\right],
\ea
\ee
where
\be
\ba
g_{+1} &= -2\sqrt{2} \left(r\mathscr{D}R_{-1}-R_{-1}\right),\\
f_{-1} &= 2\sqrt{2}ia \left[\cos\theta\,\mathscr{L}_1S_{+1}+\sin\theta\,S_{+1} \right].
\ea
\ee
Together with the expressions for $\phi_0$ and $\phi_2$, this reconstructs the full electromagnetic field associated with the chosen vacuum mode.

\subsection{Massive scalar field}

We now consider the massive scalar field on the Kerr background, which satisfies
\be
    \left(\nabla_\mu\nabla^\mu-\mu^2\right)\Phi=0,
\ee
where $\mu$ is the mass of the scalar field. This equation is separable in Boyer-Lindquist coordinates. In particular, by taking the ansatz
\be
    \Phi=e^{-i\omega t+im\varphi}S_{\ell m}^{a}(\theta) R_{\ell m}^{\, a}(r),
\ee
the Klein-Gordon equation reduces to the angular equation
\be
\ba
&\frac{1}{\sin\theta}\frac{\dif}{\dif\theta}
\left(\sin\theta\frac{\dif S_{\ell m}^{a}}{\dif\theta}\right)
+\Bigg[
a^2(\omega^2-\mu^2)\cos^2\theta
\\
&\qquad
-\frac{m^2}{\sin^2\theta}
+A_{0\ell m}
\Bigg]S_{\ell m}^{a}=0,
\ea
\ee
and the radial equation
\be
\ba\label{eq:radial0}
&\frac{\dif}{\dif r}\left(\Delta\frac{\dif R_{\ell m}^{\,a}}{\dif r}\right)
+\Bigg[
\frac{K^2}{\Delta}
-\mu^2 r^2
\\
&\qquad
-a^2\omega^2+2am\omega
-A_{0\ell m}
\Bigg]R_{\ell m}^{\,a}=0.
\ea
\ee
Similar to the massless spin-1 field, solutions to the angular equation can be given by the spin-0 spheroidal harmonic, $S_{\ell m}^a$, with spheroidicity $a\sqrt{\omega^2-\mu^2}$.

For $\omega > \mu$, the radial equation \eqref{eq:radial0} admits two linearly independent scattering solutions. The in-mode behaves as
\be
\ba
R^{\,a, \rm in}_{\omega \ell m} \to
\begin{cases}
    e^{-i k_H r_*}, \qquad &r\to r_+\\
    A^{a,\rm in}_{\omega \ell m}r^{-1}e^{-i k_\infty r_*} + A^{a,\rm out}_{\omega \ell m}r^{-1}e^{i k_\infty r_*}, \qquad &r\to\infty
\end{cases},
\ea
\ee
and the up-mode behaves as
\be
\ba
R^{\,a,\rm up}_{\omega \ell m} \to
\begin{cases}
    A^{a,\rm down}_{\omega \ell m} e^{-i k_H r_*} + A^{a,\rm up}_{\omega \ell m} e^{i k_H r_*}, \qquad &r\to r_+\\
    r^{-1}e^{i k_\infty r_*}, \qquad &r\to\infty
\end{cases},
\ea
\ee
where $k_\infty=\sqrt{\omega^2-\mu^2}$ is the asymptotic momentum. By analytically continuing $\omega$ to the complex plane, the radial equation \eqref{eq:radial0} also admits solutions with a discrete spectrum $\omega_{n\ell m}$. These solutions behave as
\be\ba\label{eq:asypq}
    R_{n \ell m}^a \to \mathsf{N}_{n \ell m}
    \begin{cases}
        A^{a,\rm down}_{n \ell m} e^{-i k_H r_*}, \qquad &r\to r_+\\
        r^{-1}e^{i k_\infty r_*}, \qquad &r\to\infty
    \end{cases} \,,
\ea\ee
where $\mathsf{N}_{n \ell m}$ is the normalization factor. In particular, solutions with ${\rm Im}\,k_\infty>0$ are known as QBSs, while solutions with ${\rm Im}\,k_\infty < 0$ are known as quasinormal modes (QNMs). Both QBSs and QNMs correspond to the poles of the radial equation's Green function because the Wronskian
\be\label{eq:Wronskian}
   \left. W[R^{a,\rm in}_{\omega \ell m }, R^{a,\rm up}_{\omega \ell m}] \right|_{\omega=\omega_{n\ell m}} = 2i k_H A^{a,\rm up}_{\omega_{n \ell m}} = 2i k_\infty A^{a,\rm in}_{\omega_{n \ell m}} = 0.
\ee

Following Refs.~\cite{Green:2022htq, Cannizzaro:2023jle}, it is useful to introduce a conserved bilinear form for the subsequent perturbation calculation. Starting from the KG equation in Kerr spacetime, one can define a ``base'' product (related to the symplectic form),
\be
    \Pi_{\mathscr{S}} [\Phi_1, \Phi_2] = \int_\Sigma \left( \Phi_1 \nabla_a \Phi_2 - \Phi_2 \nabla_a \Phi_1 \right) n^a \mathrm{d}V,
\ee
where $\mathscr{S}$ is a time slice with unit normal vector $n^a$. Gauss’s theorem allows one to verify that, if $\Phi_1, \Phi_2$ are solutions of KG equations, the product is conserved. As the Kerr spacetime is axisymmetric and stationary, we introduce the $t-\varphi$ reflection operator $\JJ$ \cite{Green:2022htq}, whose action on a scalar field simply takes $t \to -t$ and $\phi \to -\phi$. The flip operator $\JJ$ converts the symplectic form into a conserved symmetric bilinear form with respect to which the time-translation operator is symmetric, thereby ensuring the orthogonality of modes with distinct frequencies. In Boyer-Lindquist coordinates, the bilinear form is defined by 
\be
\ba
    \langle \Phi_1, \Phi_2 \rangle = & \Pi_{\mathscr{S}} [\JJ \Phi_1,\Phi_2] \\
    = & \int_{r_+}^{\infty} \dif r \int_{S^2} \dif^2 \Omega \, \Bigg[  
        \frac{2Mra}{\Delta}(\JJ\Phi_1 \partial_\varphi \Phi_2 - \Phi_2 \partial_\varphi \JJ \Phi_1) \\
        & \qquad \qquad \qquad + \frac{\Sigma}{\Delta} \left( r^2+a^2+\frac{2Mra^2}{\Sigma}\sin^2\theta \right) \\
        & \qquad \qquad \qquad \times \left( \JJ\Phi_1 \partial_t \Phi_2 - \Phi_2 \partial_t \JJ\Phi_1 \right)
    \Bigg].
\ea
\ee

With the branch cuts chosen along $(-\infty,-\mu]$ and $[\mu,+\infty)$, the continuum modes lie on the real-frequency cut. For these real-frequency modes, $\JJ$ plays the role of complex conjugation \cite{Fu:2025ztk}. With respect to this bilinear form, the QBSs $\Phi_{n\ell m}$ and continuum modes $\Phi^{\rm in/up}_{\omega\ell m}$ satisfy orthogonality relations of the form

\be\label{eq:orthogonality}
\ba
    \langle \Phi^{\rm in}_{\omega\ell m}, \Phi^{\rm in}_{\omega'\ell' m'} \rangle &= 4\pi k_\infty |A^{a,\rm in}_{\ell m}|^2 \delta(\omega-\omega')\delta_{\ell\ell'}\delta_{mm'}, \\
    \langle \Phi^{\rm up}_{\omega\ell m}, \Phi^{\rm up}_{\omega'\ell' m'} \rangle &= 4\pi k_H |A^{a,\rm up}_{\ell m}|^2 \delta(\omega-\omega')\delta_{\ell\ell'}\delta_{mm'}, \\
    \langle \Phi_{n\ell m}, \Phi_{n'\ell' m'} \rangle &= \delta_{nn'}\delta_{\ell\ell'}\delta_{mm'}.\\
\ea
\ee
Here the normalization factor in Eq.~\eqref{eq:asypq} for QBSs is given by
\be
\mathsf{N}_{n \ell m} = \left[ 2 i k_\infty A^{a,\rm down}(\omega_n) \partial_{\omega_n} A^{a,\rm in}(\omega_n) \right]^{-1/2} \,.
\ee
A derivation of this expression is provided in Appendix~\ref{app:greenfun}.

\section{Superradiance with Axion-photon coupling}
\label{sec:superradiance}

We consider axion-photon interactions mediated by the Chern-Simons coupling. The action is given by
\be
S_m=\int \dif^4 x \sqrt{-g}\, \mathcal{L},
\ee
with
\be
\mathcal{L}
=-\frac{1}{4}F_{\mu\nu}F^{\mu\nu}
-\frac{1}{2}\nabla_\mu\Phi\nabla^\mu\Phi
-\frac{1}{2}\mu^2\Phi^2
-\frac{g_{a\gamma}}{4}\Phi F_{\mu\nu}\tilde F^{\mu\nu},
\ee
where $g_{a\gamma}$ is the axion-photon coupling constant. The Lagrangian leads to the sourced Maxwell equations
\be\label{eq:Maxwell}
\nabla_\mu F^{\mu\nu} = J^\nu_{\rm eff},
\ee
with an effective current $J^\nu_{\rm eff} = -g_{a\gamma}(\nabla_\mu\Phi)\tilde F^{\mu\nu}$, and the axion equation
\be
\ba\label{eq:axion}
    \mathcal{H}_{a} \Phi =& \frac{g_{a\gamma}}{4}F_{\mu\nu}\tilde F^{\mu\nu} \\
    =& -i g_{a\gamma} (\phi_0 \phi_2 - \phi_1^2) + \mathrm{c.c.} \\
\ea
\ee
where $\mathcal{H}_{a} \equiv \nabla_\mu\nabla^\mu-\mu^2$ for simplicity. 

By projecting $J^\nu_{\rm eff}$ onto the Kinnersley tetrad and using Eq.~\eqref{eq:dualF}, one can get the Teukolsky equation in the presence of the axion-photon coupling. For the $s=-1$ Teukolsky equation, the source term in Eq.~\eqref{eq:Teukolsky} takes the form \cite{Teukolsky:1973ha}
\be\label{eq:TeukolskySource}
T_{-1} = (\Delta_{\rm NP}+\mu_{\rm NP})\rho^{-2} J_{\bar{m}} - (\delta^* + \pi - \bar{\tau}) \rho^{-2} J_n,
\ee
where
\be
\ba
\Delta_{\rm NP} &= n^{\mu}\nabla_\mu, &\rho &= -1/(r-i a \cos\theta),  \quad \mu_{\rm NP} = \rho^2 \bar{\rho}\Delta/2, \\
\delta^* &= \bar{m}^{\mu}\nabla_\mu, &\pi &=i a \rho^2 \sin \theta/\sqrt{2},\quad \tau = -i a \rho \bar{\rho} \sin \theta/\sqrt{2}
\ea
\ee
and $J_{\bar{m}}=\bar{m}_\mu J^\mu_{\rm eff}, J_n= n_\mu J^\mu_{\rm eff}$ are functions of $\Phi$ and $\phi_i$.

The total energy-momentum tensor of both axion and electromagnetic fields can be derived by varying the action with respect to the metric. It turns out that the total energy–momentum tensor can be decomposed into that of free axion and electromagnetic fields, while the Chern-Simons coupling does not contribute.
%\be
%T_{\mu\nu}
%=-\frac{2}{\sqrt{-g}}\frac{\delta S_m}{\delta g^{\mu\nu}}
%=T^{(\gamma)}_{\mu\nu}+T^{(\Phi)}_{\mu\nu}.
%\ee
It is because the coupling term $\sqrt{-g}F_{\mu\nu}\tilde F^{\mu\nu}$ is the metric-independent density corresponding to $F \wedge F$. Therefore the coupling changes the equations of motion, but does not contribute to the total energy-moment tensor.

In the following, we investigate superradiant scattering of electromagnetic waves by a Kerr black hole. Specifically, we consider a single incident wave with fixed quantum numbers $(\omega_\gamma,\ell_\gamma,m_\gamma)$, whose frequency lies in the superradiant regime $0<\omega_\gamma<m_\gamma\Omega_H$, and compute the axion-photon-coupling correction to the amplification of the reflected wave. We assume that no background axion field is present, so the axion serves only as a mediator in the scattering process. We then expand the solutions of Eqs.~\eqref{eq:Teukolsky} and \eqref{eq:axion} perturbatively in powers of $g_{a\gamma}$,
\be
\ba
\phi_{i} &= \phi_{i}^{(0)}
+g_{a\gamma}^2\phi_{i}^{(2)}
+\mathcal{O}(g_{a\gamma}^4), \quad i=0,1,2 \\
\Phi &= g_{a\gamma}\Phi^{(1)}
+\mathcal{O}(g_{a\gamma}^3).
\ea
\ee
With this expansion, the superradiant scattering amplitude and the amplification factor receive their leading corrections at order $g_{a\gamma}^2$.

\subsection{Superradiant scattering at zeroth order}

\begin{figure}
    \center
    \includegraphics[width=0.45\textwidth]{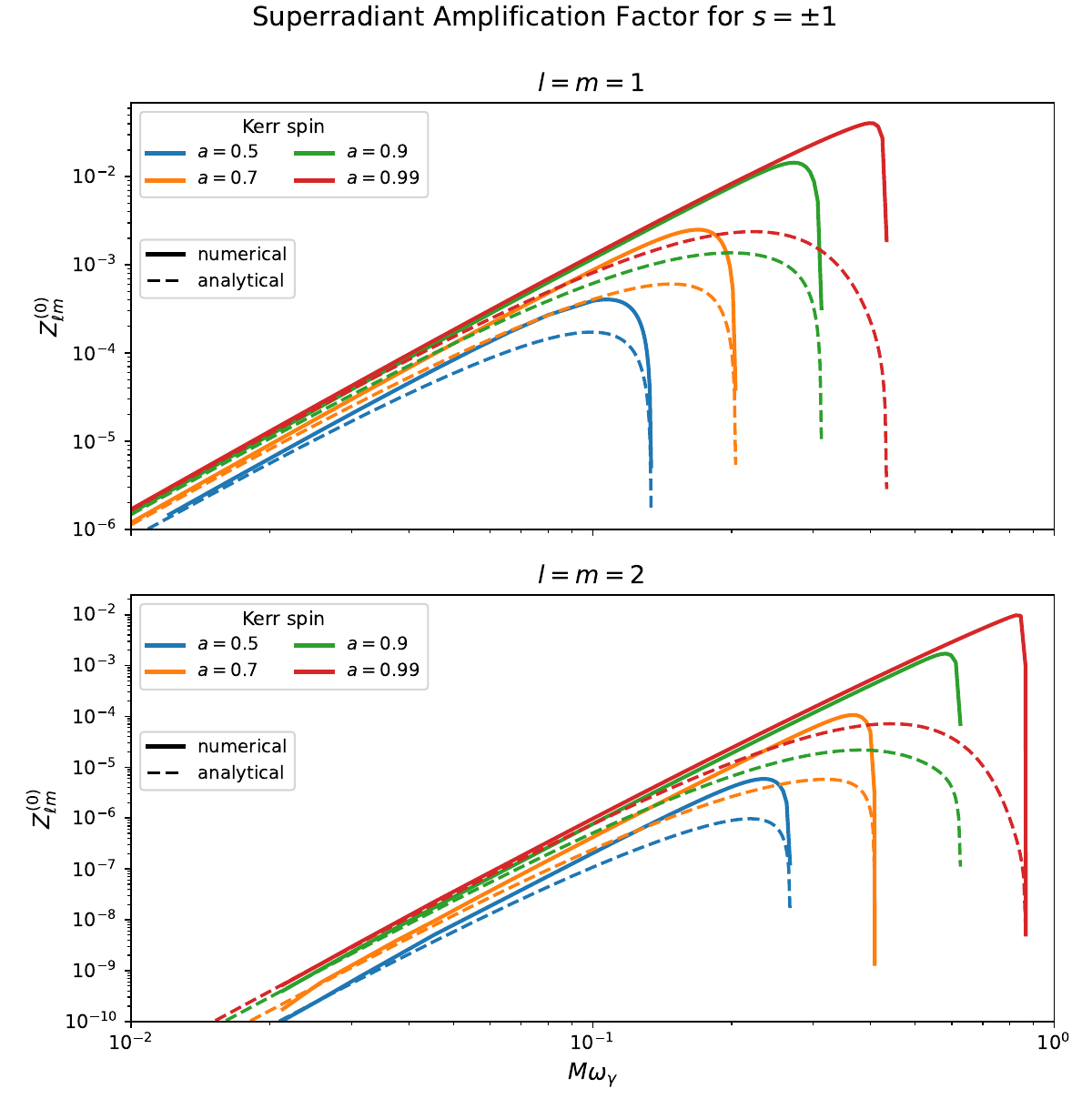}
    \caption{Zeroth-order photon superradiance amplification factor $Z^{(0)}_{\ell m}$ for the $\ell=m=1$ and $\ell=m=2$ modes at $a/M=\{0.99,0.9,0.7,0.5\}$. The solid curves denote the numerical results, obtained by solving the Teukolsky equation numerically and extracting the amplitudes from the asymptotic radial behavior. The dashed curves show the analytic low-frequency approximation.}
    \label{fig:AmplificationFactor}
\end{figure}

At zeroth order in the axion-photon coupling, the scattered electromagnetic waves obey the source-free Teukolsky equation,
\be
\mathcal{H}_{\pm1}\psi_{\pm1}^{(0)} = 0\,,
\ee
and satisfies the ingoing boundary condition at the black hole horizon. Therefore, the radial function of $\psi_{\pm1}^{(0)}$ is described by the in-mode solution. The axion field, on the other hand, is absent at zeroth order. The energy-momentum tensor is therefore purely electromagnetic. 

At spatial infinity, $\phi_0$ carries the leading incoming radiative component, whereas $\phi_2$ carries the leading outgoing radiative component \cite{Teukolsky:1974yv}. Using the asymptotic form of the in-mode solution in Eq.~\eqref{eq:innorm}, the incoming and outgoing energy fluxes are
\be\ba\label{eq:inftyflux}
    \FF_{\gamma,\rm in}^{(0)} = \lim_{r\to\infty} \int \dif \Omega\, \frac{r^2}{8\pi} |\phi_0|^2 = \frac{1}{4} |A_{1\ell m}^{\rm in}|^2 \mathcal{A}_\gamma^2, \\
    \FF_{\gamma,\rm out}^{(0)} = \lim_{r\to\infty} \int \dif \Omega\, \frac{r^2}{2\pi} |\phi_2|^2 = |A_{-1\ell m}^{\rm out}|^2 \mathcal{A}_\gamma^2.
\ea\ee
The flux at the horizon can be computed by evaluating the change in energy of the black hole. As shown in Ref.~\cite{Teukolsky:1974yv}, it is given by
\be\label{eq:BHflux}
    \FF_{\gamma,\rm BH}^{(0)} = \int \dif \Omega \frac{\omega}{k_H} 2 M r_+ T^{\mu\nu,(0)} n_\mu n_\nu = \frac{\omega}{8 M r_+ k_H} \mathcal{A}_\gamma^2 \,,
\ee
where $n_\mu$ is an inward unit vector, normal to the horizon surface. The sign of the horizon flux is determined by $k_H$. With our convention, $\FF_{\gamma,\rm BH}^{(0)}>0$ corresponds to absorption of positive energy by the black hole, while $\FF_{\gamma,\rm BH}^{(0)}<0$ corresponds to negative energy crossing the horizon and hence to energy extraction from the black hole. Energy conservation implies that
\be
    \FF_{\gamma,\rm in}^{(0)} - \FF_{\gamma,\rm out}^{(0)} = \FF_{\gamma,\rm BH}^{(0)}.
\ee
It is clear that when energy is extracted from the black hole, $k_H<0$ indicates $\FF_{\gamma,\rm in}^{(0)}<\FF_{\gamma,\rm out}^{(0)}$, and hence the amplification of electromagnetic waves. 

Finally, the superradiant amplification factor is defined by
\be
    Z^{(0)}_{s\ell m} \equiv \FF_{\gamma,\rm out}^{(0)}/\FF_{\gamma,\rm in}^{(0)}-1.
\ee
For electromagnetic waves, the flux ratio can be expressed in terms of the asymptotic amplitudes of the radial function
\be
    Z^{(0)}_{s\ell m} = \frac{\left| A^{\rm out}_{s \ell m} \right|^2}{\left| A^{\rm in}_{s \ell m} \right|^2} \left( \frac{16\omega^4}{B^2} \right)^{\pm 1} - 1,\quad \mathrm{for}\, s=\pm 1,
\ee
where the two expressions are equivalent, giving $Z^{(0)}_{+1\ell m}=Z^{(0)}_{-1\ell m}$.

In the low-frequency regime $M\omega \ll 1$, the amplification factor can be obtained analytically by using matching asymptotic techniques \cite{Starobinskii:1973vzb,Starobinskil:1974nkd,Page:1976df}. The leading result is
\be\ba\label{eq:AmpFac}
    Z_{s\ell m}^{(0)} = 4Q \left[\frac{(\ell-s)!(\ell+s)!}{(2\ell)!(2\ell+1)!!}\right]^2 \left[\omega(r_+-r_-)\right]^{2\ell+1} \prod_{k=1}^{\ell} \left(1+\frac{4Q^2}{k^2}\right)
\ea\ee
where $Q=\tfrac{r_+^2+a^2}{r_+ - r_-} (m \Omega_H-\omega)$. The expression above is only valid for $a\leq M$ and $M\omega\ll1$. Fig.~\ref{fig:AmplificationFactor} compares the analytical amplification factor given in Eq.~\eqref{eq:AmpFac} with the numerical results as a function of the dimensionless frequency $M\omega_\gamma$.
It shows explicitly that the reflected wave is amplified in the superradiant regime $0<\omega<m\Omega_H$ for which $Z_{s\ell m}^{(0)}>0$. The overall factor $\omega^{2\ell+1}$ reflects the low-frequency suppression of the wave component transmitted through the black-hole potential barrier.

\subsection{Leading axion-induced corrections}

At first order in the axion-photon coupling, the zeroth-order electromagnetic field acts as a source for the axion,
\be\label{eq:axion-source}
    \mathcal{H}_{a}\Phi^{(1)}
    =-i \mathcal{A}_\gamma^2 \left(\hat{\phi}_0^{(0)} \hat{\phi}_2^{(0)} - (\hat{\phi}^{(0)}_1)^2\right) + \mathrm{c.c.}.
\ee
Because $\Phi^{(1)}$ is proportional to $\mathcal{A}_\gamma^2$, we define $\hat{\Phi}^{(1)} = \Phi^{(1)}/\mathcal{A}_\gamma^2$ to extract the dependence on the photon amplitude. 

For the single monochromatic in-mode whose radial function is given by Eq.~\eqref{eq:innorm}, the NP scalars $\phi_i^{(0)}$ denote the positive-frequency components of electromagnetic field, while their complex conjugates $(\phi_i^{(0)})^*$ are the corresponding negative-frequency components. Since the axion source is proportional to $\phi_0^{(0)}\phi_2^{(0)}-(\phi_1^{(0)})^2$, the positive-frequency part of the source contains only $(\omega_a,m_a)=(2\omega_\gamma,2m_\gamma)$ component together with its complex conjugate. Thus, in the single-mode setup, the axion-photon interaction generates only the $s$-channel intermediate axion and the $t$- and $u$-channel contributions are absent. The $t$- and $u$-channel contributions require a helicity flip between the incoming and outgoing photons and therefore involve both photon helicities. They are consequently absent in the single-helicity sector considered here.

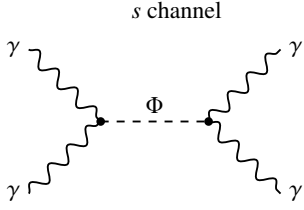
\begin{figure}
\centering
\begin{tikzpicture}[line width=0.7pt, scale=0.95]
    \node at (1.05,1.55) {$s$ channel};
    \coordinate (sL) at (0,0);
    \coordinate (sR) at (1.5,0);
    \draw[snake it] (-1,) -- (sL);
    \draw[snake it] (-1,-1) -- (sL);
    \draw[dashed] (sL) -- node[above] {$\Phi$} (sR);
    \draw[snake it] (sR) -- (2.5,1);
    \draw[snake it] (sR) -- (2.5,-1);
    \fill (sL) circle (1.7pt);
    \fill (sR) circle (1.7pt);
    \node[left] at (-1,1) {$\gamma$};
    \node[left] at (-1,-1) {$\gamma$};
    \node[right] at (2.5,1) {$\gamma$};
    \node[right] at (2.5,-1) {$\gamma$};
\end{tikzpicture}

\caption{Schematic representation of the $s$-channel axion-mediated correction to photon scattering at leading order in the coupling.}
\label{fig:st-channel}
\end{figure}

The generated axion field feeds back into the Maxwell equations through the effective current $J_{\rm eff}^{\nu}$. For the $s$-channel axion mode constructed above, $\Phi^{(1)}$ has $(\omega_a,m_a)=(2\omega_\gamma,2m_\gamma)$. To source the correction to the original photon mode, frequency matching requires the electromagnetic factor in $J_{\rm eff}^{\nu}$ to be the negative-frequency component of the zeroth-order field, $J_{s,\rm eff}^{\nu} =-(\nabla_\mu\Phi^{(1)})\tilde F^{(0)\mu\nu}_{-}[\phi_i^*]$. Hence
\be
\ba
    \mathcal{H}_{-1} \hat{\psi}_{-1}^{(2)} = 4 \pi \Sigma T_{-1}[\hat{\Phi}^{(1)},\hat{\phi}_i^*].
\ea
\ee
The sourced Teukolsky equation determines the correction to the photon wave function. We can factor out the dependence on the incident photon amplitude by defining $\hat{\psi}_{-1}^{(2)}=\psi_{-1}/\mathcal{A}_\gamma^2$.

We now turn to the energy-momentum tensor at leading order in the coupling. Since the Chern-Simons interaction does not contribute to the energy-momentum tensor, the order-$g_{a\gamma}^2$ correction to the total tensor can be decomposed into the axion contribution and the photon correction,
\be
    T^{\mu\nu,(2)}=T_{a}^{\mu\nu,(2)}+T_{\gamma}^{\mu\nu,(2)}.
\ee
This is because the axion flux is quadratic in the first-order axion field $T_{a}^{\mu\nu,(2)} \sim (g_{a\gamma}\Phi^{(1)})^2$, while the photon flux correction is linear in the second-order photon perturbation $T_{\gamma}^{\mu\nu,(2)} \sim g_{a\gamma}^2 \phi_i^{(2)} \phi_i^{(0)}$. The stationary energy flux balance at order-$g_{a\gamma}^2$ reads
\be
\FF_{\gamma,\rm out}^{(2)}+\FF_{a,\rm out}^{(2)}+\FF_{\gamma,\rm BH}^{(2)}+\FF_{a,\rm BH}^{(2)} = 0.
\ee
The axion fluxes take the form
\be\ba\label{eq:axionflux}
    &\FF^{(2)}_{a,\rm out} = 2 \omega_a k_\infty |\mathcal{A}_{a,\infty}|^2, \\
    &\FF^{(2)}_{a,\rm BH} = M r_+ \omega_a (\omega_a-m_a \Omega_H) |\mathcal{A}_{a,\rm BH}|^2,
\ea\ee
where $\mathcal{A}_{a,\infty}$ and $\mathcal{A}_{a,\rm BH}$ denote the asymptotic amplitudes of the excited axion field at spatial infinity and the black-hole horizon, respectively. Further, the incoming photon amplitude is kept fixed, and no incoming axion radiation is imposed.

The correction to the amplification factor at ${\cal O}\,(g_{a\gamma}^2)$ is defined by
\be
    \delta Z_\gamma = \FF_{\gamma,\rm out}^{(2)} / \FF_{\gamma,\rm in}^{(0)}.
\ee

To quantify the relative size of the axion-induced correction, we compare it with the zeroth-order photon amplification factor. We define
\be
    \eta \equiv \delta Z_\gamma / Z^{(0)}_{s \ell m} 
\ee
Thus $|\eta|\ll1$ corresponds to a small perturbative correction, and the sign of $\eta$ indicates whether the coupling enhances ($\eta>0$) or suppresses ($\eta<0$) the zeroth-order amplification.

In the following sections, we evaluate the correction in two physically distinct parameter regimes, distinguished by whether the sourced axion frequency lies above or below the mass threshold $\mu$, namely whether the axion can propagate to spatial infinity. To understand the dominatant physical mechanisms in the two parameter regimes, we focus on two limits $\omega_a\gg\mu$ and $\omega_a\ll\mu$. In the former limit, $k_\infty\simeq\omega_a$ is real, and the sourced axion can propagate efficiently to spatial infinity and carry away energy. In the latter limit, the axion field is confined by the mass barrier and the correction is governed by the energy exchange with the black hole horizon. Moreover, we consistently assume the low-frequency conditions $M\omega_\gamma\ll1$ and $M\mu\ll1$ throughout our analysis of both regimes.

\subsection{Radiative limit \texorpdfstring{$\omega_a \gg \mu$}{omega >> mu}}

\begin{figure}[b]
    \centering
    \includegraphics[width=0.45\textwidth]{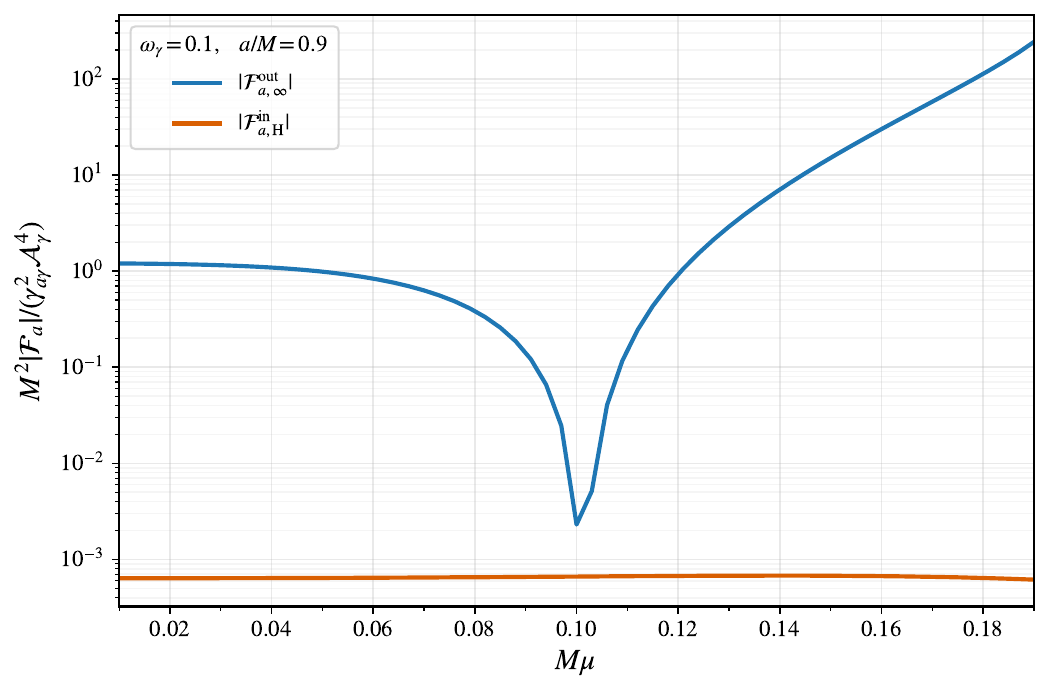}
    \caption{Comparison of the outgoing axion flux at spatial infinity with the horizon fluxes for axions produced through the $s$-channel process, for $a/M=0.9$ and $\omega_\gamma=0.1$.     }
    \label{fig:FluxComparison}
\end{figure}

We first consider the radiative limit, in which the axion produced by the electromagnetic source has $\omega_a=2\omega_\gamma \gg \mu$ and can propagate to spatial infinity. 

The characteristic height of the effective potential is set by the black hole scale, $V_{\rm eff} \sim M^{-2}$ \cite{Chandrasekhar:1985kt}. Since the effective photon energy is of order $\omega_\gamma^2$, the low-frequency condition $M\omega_\gamma\ll1$ implies $V_{\rm eff}/\omega_\gamma^2\sim(M\omega_\gamma)^{-2}\gg1$. Transmission into the near-horizon region is therefore strongly suppressed, and the electromagnetic field is predominantly supported outside the potential barrier.

The source term in Eq.~\eqref{eq:axion-source}, and hence the axion generated by interaction, is therefore also predominantly located outside the potential barrier. Since $M\omega_a\ll1$, the inward-propagating axion component encounters a similarly high effective potential barrier, and its transmission into the near-horizon region is strongly suppressed. By contrast, when $\omega_a\gg\mu$, the outward-propagating axion component can reach spatial infinity and produce a non-zero outgoing energy flux that is much larger than the horizon flux. The axion fluxes shown in Fig.~\ref{fig:FluxComparison} are evaluated according to Eq.~\eqref{eq:axionflux}, with the asymptotic amplitudes $\mathcal{A}_{a,\infty}$ and $\mathcal{A}_{a,\rm BH}$ obtained from the inhomogeneous axion solution using the Green's-function method described below. The outgoing flux exhibits a dip near $\mu\simeq\omega_\gamma$, where the overlap between the source and the axion wave function is strongly suppressed. As $M\mu$ approaches $M\omega_a=0.2$, the momentum  $k_\infty=\sqrt{\omega_a^2-\mu^2}$ tends to zero, while the overlap between the axion mode and the electromagnetic source is enhanced.

We therefore obtain
\be
\left|\FF_{\gamma,\rm BH}^{(2)}\right|
+\left|\FF_{a,\rm BH}^{(2)}\right|
\ll
\left|\FF_{\gamma,\rm out}^{(2)}\right|
+\left|\FF_{a,\rm out}^{(2)}\right| .
\ee
The conservation of the order-$g_{a\gamma}^2$ flux then reduces to
\be
\FF_{\gamma,\rm out}^{(2)} \approx -\FF_{a,\rm out}^{(2)} .
\ee
Thus, the correction to the outgoing photon flux can be obtained directly from the outgoing axion flux. This gives the approximate correction to the photon amplification factor,
\be
    \delta Z_\gamma \approx
    - \FF_{a,\rm out}^{(2)} / \FF_{\gamma,\rm in}^{(0)} .
\ee

The outgoing axion flux $\FF_{a,\rm out}^{(2)}$ can be obtained by solving the inhomogeneous axion equation using the Green's function $G_a(x,x')$, which satisfies
\be
\mathcal{H}_{a} G_a(x,x') = \frac{\delta^{(4)}(x-x')}{\sqrt{-g}}.
\ee
Then the inhomogeneous solution of Eq.~\eqref{eq:axion-source} can be expressed as
\be\label{eq:AxionGreen}
    \hat{\Phi}^{(1)} = \int \dif^4 x'\sqrt{-g}\, G_{a}(x,x') \mathcal{S}_{a}(r',\theta') e^{-i \omega_a t' + i m_a \varphi'} + {\rm c.c.},
\ee
where we define $\mathcal{S}_{a}(r,\theta)e^{-i \omega_a t + i m_a \varphi} \equiv -i \left( \hat{\phi}_0^{(0)} \hat{\phi}_2^{(0)} - (\hat{\phi}_1^{(0)})^2 \right)$ for simplicity.

Since the Kerr background is stationary and axisymmetric, the Green's function can be decomposed into frequency and azimuthal modes as
\be
\ba
    G_a(x,x') = \sum_{\ell m} \int & \frac{\dif\omega}{2\pi} e^{-i\omega(t-t')+im(\varphi-\varphi')} \\
    & \times S^a_{\ell m,\omega}(\theta) S^a_{\ell m,\omega}(\theta') G_{a,\omega\ell m}(r,r'),
\ea
\ee
The radial Green's function is constructed from two independent homogeneous solutions of the axion radial equation,
\be
\ba
    G_{a,\omega\ell m}(r,r') = \frac{R^{a,\rm in}_{\omega\ell m}(r_<) R^{a,\rm up}_{\omega\ell m}(r_>)}{W[R^{a,\rm in}_{\omega\ell m},R^{a,\rm up}_{\omega\ell m}]},
\ea
\ee
where $r_<=\min(r,r')$ and $r_>=\max(r,r')$.

The angular part of the source term can be expanded in the spin-0 spheroidal harmonics
\be\label{eq:SaExpand}
    \mathcal{S}_{a}(r,\theta)= \sum_{\ell \ge \abs{m_a}} C_{\ell}(r) S^{a}_{\ell m_a,\omega_a}(\theta),
\ee
where $C_{\ell}(r)$ are projection coefficients. 

In the limit $a\omega_a \ll 1$, the spin-0 spheroidal harmonics reduce to spherical harmonics $Y_{\ell m}$. At this order, the angular selection rules select the multipole $\ell_a=2\ell_\gamma$. Higher multipoles with $\ell_a>2\ell_\gamma$ arise only from spheroidal harmonic mixing and are suppressed by powers of $a\omega_a$; we neglect them in what follows.

Since the outgoing flux is evaluated at spatial infinity, the observation point lies outside the source region. Thus, for the radial Green's function we have $r>r'$ over the support of the source. The asymptotic field is therefore proportional to $R^{a,\rm up}(r)$, while the source overlap involves only $R^{a,\rm in}(r')$. The axion field can be written as
\be
    \hat{\Phi}^{(1)} = \mathcal{A}_a e^{-i \omega_a t + i m_a \varphi} S^{a}_{\ell_a m_a}(\theta) R^{a,\rm up}_{\ell_a m_a,\omega_a}(r) + \rm{c.c.}
\ee
with
\be\label{eq:axionamp}
\ba
    \mathcal{A}_a \equiv & \frac{2 \pi}{W_a} \\
    & \times \int_{r_+}^{\infty} \dif r' \int_0^\pi \dif\theta' R^{a,\rm in}_{\ell_a m_a,\omega_a}(r') S^{a}_{\ell_a m_a,\omega_a}(\theta') \mathcal{S}_{a}(r',\theta') \Sigma \sin \theta'
\ea
\ee
according to Eq.~\eqref{eq:AxionGreen}. When ${\rm Im}\,\omega_{n\ell m}<0$, the axion radial function diverges near the horizon, which makes the numerical evaluation of the overlap integrals in Eq.~\eqref{eq:axionamp} ill-behaved. We therefore introduce a regularization method in Appendix~\ref{app:numerical-method}, following Ref.~\cite{Cannizzaro:2023jle}, in which the divergent boundary terms are subtracted explicitly. This procedure can be extended straightforwardly to the electromagnetic case, as well as to divergences at spatial infinity. The outgoing axion flux is then
\be
    \FF_{a,\rm out}^{(2)} = 2 \omega_a k_\infty g_{a\gamma}^2  \mathcal{A}_\gamma^4 |\mathcal{A}_a|^2.
\ee

\begin{figure}
    \centering
    \includegraphics[width=0.45\textwidth]{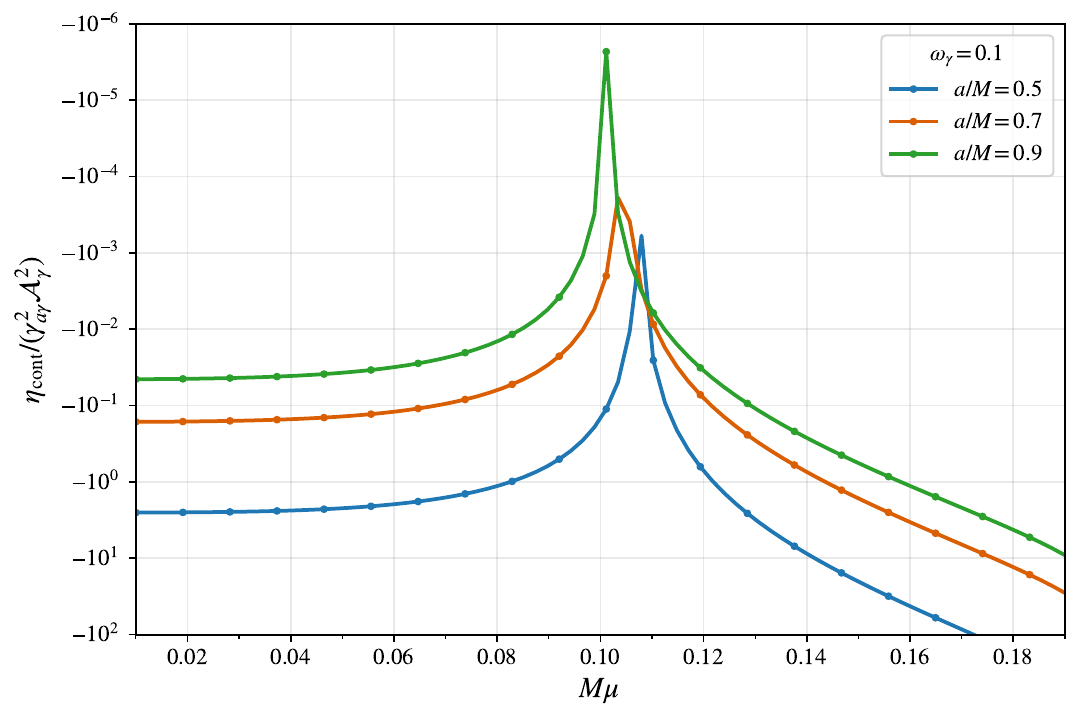}
    \caption{The superradiance correction contributed by outgoing axion flux in the radiative regime, shown as a function of $M \mu$ for black hole spins $a/M=0.5,0.7,0.9$ and photon energy $\omega_\gamma=0.1$. The cutoff of the flux at $\mu=2\omega_\gamma$ is due to the mass threshold.}
    \label{fig:InfinityEmit}
\end{figure}

Numerical results for the outgoing axion flux are shown in Fig.~\ref{fig:InfinityEmit}. Here we extract the dependence of $g_{a\gamma}^2 \mathcal{A}_\gamma^2$, since one can choose a small coupling constant $g_{a\gamma}$ to satisfy the perturbative condition. Since the correction closely tracks the axion flux at infinity, it likewise exhibits a dip near $\mu\simeq\omega_\gamma$ and an enhancement as $\mu\to2\omega_\gamma$. The vanishing point shifts slightly with the black hole spin because the spin modifies the radial profiles of the relevant axion and photon modes.

\subsection{Confined limit \texorpdfstring{$\omega_a \ll \mu$}{omega << mu}}

We next consider the confined limit, in which the sourced axion frequency lies below the mass threshold, $\omega_a=2\omega_\gamma\ll\mu$. In this case, the radial momentum at infinity $k_\infty=\sqrt{\omega_a^2-\mu^2}=i\kappa$ is purely imaginary, where $\kappa>0$ follows from the bound-state Riemann sheet chosen above. Consequently, the axion solution is evanescent, $\Phi\sim e^{-\kappa r_*}$, and the outgoing axion flux vanishes. Hence the photon flux correction at ${\cal O} (g_{a\gamma}^2)$ is fixed by the horizon fluxes, 
\be
    \FF_{\gamma,\rm out}^{(2)} \approx -\left( \FF_{\gamma,\rm BH}^{(2)} + \FF_{a,\rm BH}^{(2)} \right).
\ee

In this regime, the outgoing axion flux at infinity vanishes. The outgoing photon flux correction therefore cannot be inferred from axion emission alone, as in the radiative regime. Instead, we compute it directly from the second-order photon perturbation by solving the coupled system
\be
\ba
\mathcal{H}_{a}\hat{\Phi}^{(1)} &= \mathcal{S}_{a}[\hat{\phi}_i^{(0)}] e^{-i \omega_a t + i m_a \varphi},\\
\mathcal{H}_{-1}\hat{\psi}_{-1}^{(2)} &= \mathcal{S}_{\gamma}[\hat{\Phi}^{(1)},\hat{\phi}_i^{(0)*}] e^{-i \omega_\gamma t + i m_\gamma \varphi},
\ea
\ee
where $\mathcal{S}_{\gamma}(r,\theta) e^{-i \omega_\gamma t + i m_\gamma \varphi} \equiv 4 \pi \Sigma T_{-1}[J^\nu_{s,\rm eff}]$ denotes the photon source term corresponding to the $s$-channel.

The solution of the inhomogeneous axion equation can be decomposed into the discrete QBS spectrum and the continuum spectrum, as proved in Appendix~\ref{app:greenfun}. The discrete part is spanned by the QBS radial functions $R^a_{n\ell m}$, while the continuum part is spanned by the scattering in-modes. We write
\be\label{eq:AxionDecomp}
\ba
    \hat{\Phi}^{(1)} = & \sum_{\ell m} S^a_{\ell m}(\theta) e^{-i \omega_a t + i m_a \varphi} \\
    & \quad \times \left( \sum_n c^a_{n \ell m} R^{a}_{n \ell m}(r) + \int_\mu^\infty \dif\omega' c^a_{\omega' \ell m} R^{a,\rm in}_{\omega' \ell m}(r) \right) + \rm{c.c.},
\ea
\ee
with
\be
\ba
    c^a_{n \ell m} &= \frac{2\pi}{-i(\omega_a - \omega_{n \ell m})} \int \dif r' \dif\theta' R^{a}_{n \ell m}(r') S^a_{\ell m}(\theta') \mathcal{S}_a(r',\theta') \Sigma \sin\theta',\\
    c^a_{\omega' \ell m} &= \frac{2\pi}{-i(\omega_a - \omega')} \int \dif r' \dif\theta' \frac{R^{a,\rm in}_{\omega' \ell m}(r') S^a_{\ell m}(\theta') \mathcal{S}_a(r',\theta')}{2 i k_\infty A^{a,\rm in}_{\omega'} A^{a,\rm out}_{\omega'}} \Sigma \sin\theta'.   
\ea
\ee

Substituting the axion solution $\hat{\Phi}^{(1)}$ into the sourced photon equation, we solve for the order-$g_{a\gamma}^2$ correction to the photon field using the Green's function method.
\be
    \hat{\psi}_{-1}^{(2)}|_{r \to \infty} = \delta \mathcal{A}^{(2)}_\gamma \, R^{\rm up}_{-1 \ell m}(r) S_{-1 \ell m}(\theta) e^{-i \omega_\gamma t + i m_\gamma \varphi} + \rm{c.c.},
\ee
where
\be
\ba
    \delta \mathcal{A}^{(2)}_\gamma = & \frac{2 \pi}{W_{-1}} \\
    & \times \int_{r_+}^{\infty} \dif r' \int_0^\pi \dif\theta' R^{\rm in}_{-1}(r') S_{-1 \ell m}(\theta') \mathcal{S}_\gamma[\hat{\Phi}^{(1)},\hat{\phi}_i^{*(0)}] \Sigma \sin\theta'
\ea
\ee
and $W_{-1} = R^{\rm in}_{-1\ell m} \partial_r R^{\rm up}_{-1\ell m} - R^{\rm up}_{-1\ell m} \partial_r R^{\rm in}_{-1\ell m}$ is a constant.

Since $\mathcal{S}_{\gamma}$ is linear in $\Phi^{(1)}$, the photon response can be decomposed into the same axion eigenmodes. Substituting the expansion of
$\Phi^{(1)}$ into the source gives
\be
    \delta\mathcal{A}^{(2)}_\gamma = \sum_{\ell m}\left[
    \sum_n c^a_{n\ell m} c_{n\ell m}^{\gamma}
    +\int \dif\omega' c^a_{\omega'\ell m} c_{\omega'\ell m}^{\gamma}
    \right],
\ee
where the photon overlap integrals contributed by discrete modes and continuous modes are
\be
\ba
    c_{n\ell m}^{\gamma} &= \frac{2 \pi}{W_{-1}} \int_{r_+}^{\infty}\dif r'\int_0^\pi \dif\theta'R^{\rm in}_{-1}(r')S_{-1\ell_\gamma m_\gamma}(\theta') \\
    & \qquad \qquad \qquad \times \mathcal{S}_\gamma[\hat{u}_{n\ell m},\hat{\phi}_i^{*(0)}]\Sigma\sin\theta',\\ 
    c_{\omega'\ell m}^{\gamma} &= \frac{2 \pi}{W_{-1}} \int_{r_+}^{\infty}\dif r'\int_0^\pi \dif\theta' R^{\rm in}_{-1}(r')S_{-1\ell_\gamma m_\gamma}(\theta') \\
    & \qquad \qquad \qquad \times \mathcal{S}_\gamma[\hat{u}_{\omega'\ell m},\hat{\phi}_i^{*(0)}]\Sigma\sin\theta'.
\ea
\ee
Here
\be
\ba
    \hat{u}_{n\ell m} &= e^{-i\omega_a t+im_a\varphi} S^a_{\ell m}(\theta)R^{a}_{n\ell m}(r),\\
    \hat{u}_{\omega'\ell m} &= e^{-i\omega_a t+im_a\varphi} S^a_{\ell m}(\theta)R^{a,\rm in}_{\omega'\ell m}(r).
\ea
\ee
Thus the coefficients $c^a_{n\ell m}$ and $c^a_{\omega'\ell m}$ encode how strongly
the electromagnetic background excites each axion eigenmode, while $c_{n\ell m}^{\gamma}$ and $c_{\omega'\ell m}^{\gamma}$ describe how each axion mode sources the outgoing photon correction.

In this case, the modified amplification factor can be expressed as
\be
\ba
    Z_{s\ell m} &= \frac{\left|\mathcal{A}_\gamma A^{\rm out}_{-1 \ell m} + g_{a\gamma}^2 \mathcal{A}_\gamma^3 \delta \mathcal{A}^{(2)}_\gamma \right|^2}{\left| \mathcal{A}_\gamma A^{\rm in}_{-1 \ell m} \right|^2} \left( \frac{16\omega^4}{B^2} \right)^{-1} - 1 \\
    &= Z_{s\ell m}^{(0)} + 2 g_{a\gamma}^2 \mathcal{A}_\gamma^2 \frac{\Re\, \left\{ (A_{-1 \ell m}^{\rm out})^* \delta \mathcal{A}_\gamma^{(2)}\right\}}{\left|A^{\rm in}_{-1 \ell m}\right|^2} \left( \frac{16\omega^4}{B^2} \right)^{-1} + \mathcal{O}(g_{a\gamma}^4).
\ea
\ee
Hence the correction to the amplification factor is
\be
    \delta Z_\gamma = 2 g_{a\gamma}^2 \mathcal{A}_\gamma^2 \frac{\Re\, \left\{ (A_{-1 \ell m}^{\rm out})^* \delta \mathcal{A}_\gamma^{(2)}\right\}}{\left|A^{\rm in}_{-1 \ell m}\right|^2} \left( \frac{16\omega^4}{B^2} \right)^{-1}.
\ee

\begin{figure}
    \centering
    \includegraphics[width=0.45\textwidth]{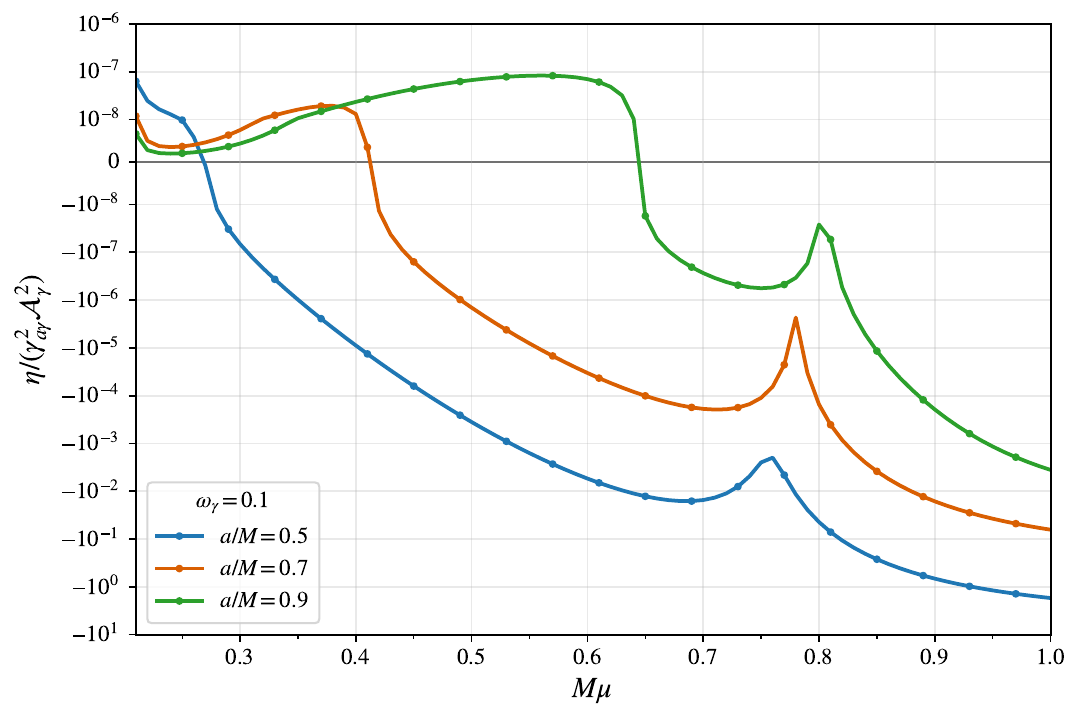}
    \caption{The superradiance correction contributed by QBSs in the confined regime, shown as a function of $M \mu$ for black hole spins $a/M=0.5,0.7,0.9$ and photon energy $\omega_\gamma=0.1$. Positive and negative values correspond to enhancement and suppression of the photon amplification, respectively. The sign changes occur when the relevant QBS crosses the superradiant threshold.}
    \label{fig:Bound} 
\end{figure}

In flat spacetime, the axion and photon equations are Hermitian. The induced amplitude correction to the outgoing photon amplitude is therefore purely dispersive: it changes only the phase of the outgoing wave and does not modify its flux. This implies
\be
    \Re\,(A_{-1 \ell m}^{\rm out})^* \delta \mathcal{A}_\gamma^{(2)}=0.
\ee
In the Kerr background, the presence of the horizon makes the problem effectively non-Hermitian. As a result, the correction $\delta\mathcal{A}_\gamma$ can also change the norm of the outgoing photon amplitude. The non-Hermitian contribution to $\delta\mathcal{A}_\gamma$ has two origins: the imaginary part of the axion eigenfrequencies, ${\rm Im}\,\omega_{n\ell m}$, and the non-Hermitian part of the wave-function overlap integrals. Since the electromagnetic source is mainly supported far from the black hole, while the non-Hermitian part of the wave functions is localized near the horizon, their overlap is small. An order-of-magnitude estimate of the non-Hermitian contributions from the complex eigenfrequency and the overlap integrals is given in Appendix~\ref{app:nonHerm}. For the modes considered here, $\ell_a \geq 2$, the overlap-integral contribution is more strongly suppressed in the $M\mu\ll1$ limit. Therefore, the dominant non-Hermitian effect is expected to be carried by ${\rm Im}\,\omega_{n\ell m}$. Keeping only this leading contribution, we approximate the amplitude correction as
\be 
\ba
    \delta \mathcal{A}_\gamma \simeq & \frac{4\pi^2 i}{W_{-1}} \sum_{n \ell m} {\rm \frac{\mathcal{I}_{n \ell m}}{\omega_a-\omega_{n \ell m}}} \\
\ea
\ee
with
\be
\ba
    \mathcal{I}_{n \ell m} = & \left( \int \dif r' \dif\theta' R^a_{n \ell m}(r') S^a_{\ell m}(\theta') \mathcal{S}_{a}(r',\theta') \Sigma \sin\theta' \right) \\
    & \times \left( \int \dif r' \dif\theta' R^{\rm in}_{-1}(r')S_{-1\ell_\gamma m_\gamma}(\theta') \mathcal{S}_\gamma[\hat{u}_{n\ell m}]\Sigma\sin\theta' \right)
\ea
\ee
and we shall neglect the non-Hermitian part of the overlap integral for numerical simplicity. The numerical results are shown in Fig.~\ref{fig:Bound}. Since the correction is controlled primarily by the imaginary part of the QBS frequencies, its sign reflects the direction of energy exchange between the QBSs and the black hole. At small $M\mu$, QBSs satisfy the superradiant condition and extract energy from the black hole, resulting in a positive correction to the photon amplification. As $M\mu$ increases, the QBS frequencies cross the superradiant bound, beyond which the modes are absorbed by the horizon and the correction becomes negative. The corresponding sign-changing point shifts as the black hole spin increases. The additional near-zero feature also arises from the vanishing of the overlap integral $\mathcal{I}_{n \ell m}$.

\section{Conclusion}
\label{sec:conclusion}

In this work we studied the effect of axion-photon coupling on photon superradiant scattering in Kerr spacetime. Treating the electromagnetic and axion fields as test fields on a fixed Kerr background, we formulated the coupled system in a fully relativistic framework and solved it perturbatively in the coupling $g_{a\gamma}$. Within this framework, we clarified how the correction to the photon superradiant amplification factor is governed by the axion energy fluxes through spatial infinity and the black hole horizon.

For a monochromatic photon incident from spatial infinity, the Chern-Simons coupling induces an $s$-channel correction for electromagnetic scattering, mediated by an axion propagator with $(\omega_a,m_a)=(2\omega_\gamma,2m_\gamma)$. Whether the axion frequency lies above or below the mass threshold determines the dominant channel of energy transfer. Above the threshold, the axion can propagate to spatial infinity and carry energy away, whereas below the threshold it is confined and exchanges energy with the black hole through the horizon. The corresponding axion fluxes determine the correction to the photon superradiant amplification factor, whose sign indicates whether the axion-mediated interaction removes energy from the photon wave or enhances it with energy extracted from the black hole.

In the radiative limit, $\omega_a \gg \mu$, the sourced axion has a propagating component at spatial infinity and carries energy away. Under the low-frequency condition $M\omega_\gamma\ll1$, the electromagnetic field and the resulting axion source are predominantly supported outside the black-hole potential barrier. Transmission toward the horizon is therefore suppressed, and the correction is governed primarily by the outgoing axion flux at infinity.

In the confined limit, $\omega_a \ll \mu$, the axion field is evanescent at spatial infinity and carries no outgoing flux. The correction then depends on the horizon flux, which is dominated by the imaginary parts of the QBS frequencies. When the relevant QBSs satisfy the superradiant condition ${\rm Re}\,\omega_{n\ell m}<m\Omega_H$, they extract energy from the black hole and produce a positive correction to photon amplification. Conversely, when the superradiant condition is not satisfied, the QBSs lose energy through the horizon and produce a negative correction.

\section*{ACKNOWLEDGMENTS}
J. Z. is supported by the National Natural Science Foundation of China (NSFC) under Grants No.~E414660101 and No.~12347103, the Fundamental Research Funds for the Central Universities under Grants No.~E4EQ6604X2 and No.~E3ER6601A2, and the National Key R\&D Program of China, No.~2025YFE0217300.

\appendix

\section{Numerical Method}
\label{app:numerical-method}

In the numerical calculation, the radial integral $\int_{r_+}^{\infty}dr\,\mathcal{K}(r)$ can become ill behaved at the horizon. For example, a QBS radial function behaves as
\be
    R_{\rm QBS} \sim e^{-i k_H r_*} \sim (r-r_+)^{r_+ {\rm Im}\, \omega_n}
\ee
near horizon. For ${\rm Im}\,\omega_n<0$, this factor can grow toward the horizon. Additional powers of $\Delta^n \sim (r-r_+)^n$ with $n \leq 0$ can further enhance the endpoint singularity. In this appendix, we introduce a counterterm method to regularize these divergences \cite{Cannizzaro:2023jle}.

In the near-horizon limit $r \rightarrow r_+$, the integrand can be expanded as
\be
    \mathcal{K}(r) = \sum_\alpha \mathcal{K}^{\rm reg,H}_\alpha (r-r_+)^{p_\alpha+i q_\alpha}
\ee
where $\mathcal{K}^{\rm reg,H}_\alpha$ denotes the regular part of the
integrand, which approaches a constant near the horizon, and $q_\alpha$
describes the oscillatory phase. For $p_\alpha \leq 0$, the corresponding
branch is singular at the endpoint. We therefore introduce a cutoff
$r=r_+ + \epsilon$ with $\epsilon \ll 1$ and separate the integral into two
pieces,
\be  
    \int_{r_+}^\infty \dif r\,\mathcal{K}(r)
    =
    \int_{r_+}^{r_+ + \epsilon} \dif r\,\mathcal{K}(r)
    +
    \int_{r_+ + \epsilon}^\infty \dif r\,\mathcal{K}(r).
\ee

In the interval $(r_+,r_++\epsilon)$, the integrand is approximated by its
singular asymptotic branches. The corresponding primitive is
\be
    \mathcal{P}_{\rm H}(r)
    =
    \sum_\alpha \mathcal{K}^{\rm reg,H}_\alpha
    \begin{cases}
    \dfrac{(r-r_+)^{p_\alpha+1+i q_\alpha}}
    {p_\alpha+1+i q_\alpha},
    & p_\alpha+1+i q_\alpha \neq 0, \\[1.2ex]
    \log(r-r_+),
    & p_\alpha+1+i q_\alpha = 0 .
    \end{cases}
\ee
where the sum runs over the singular near-horizon branches. The contribution
from the near-horizon region is then
\be
    \int_{r_+}^{r_+ + \epsilon} \dif r\,\mathcal{K}(r)
    \simeq
    \mathcal{P}_{\rm H}(r_+ + \epsilon) - \mathcal{P}_{\rm H}(r_+)
    ,
\ee
where the divergent part of $\mathcal{P}_{\rm H}(r_+)$ is discarded. After
this subtraction, the regularized integral gives a finite numerical result
which is independent of the cutoff parameter $\epsilon$.

The same idea can be applied to the large-radius tail. In the far region, the
integrand is expanded as
\be
    \mathcal{K}(r)
    =
    \sum_\beta \mathcal{K}_{\beta}^{\rm reg,\infty}\,
    r^{p_\beta+i q_\beta}
    e^{-\kappa_\beta r}
    e^{i k_\beta r},
    \qquad r\rightarrow\infty .
\ee
Here $\kappa_\beta>0$ for bound-state integrals, while $\kappa_\beta=0$ for
continuum integrals. The primitive of the asymptotic tail is
\be
    \mathcal{P}_{\infty}(r)
    =
    -\sum_\beta
    \mathcal{K}_{\beta}^{\rm reg,\infty}
    \lambda_\beta^{-\nu_\beta-1}
    \Gamma(\nu_\beta+1,\lambda_\beta r),
\ee
where the sum runs over the far-zone divergent branches, with $\nu_\beta=p_\beta+i q_\beta$ and $\lambda_\beta=\kappa_\beta-i k_\beta$. The tail contribution is therefore included through
\be
\ba
    \int_{r_++\epsilon}^{\infty} \dif r\,\mathcal{K}(r)
    &\simeq
    \int_{r_++\epsilon}^{r_{\rm max}} \dif r\,\mathcal{K}(r)
    - \mathcal{P}_{\infty}(r_{\rm max}).
\ea
\ee
Here $\Gamma(a,z)$ is the upper incomplete Gamma function.

\section{Retarded Green's function and mode decomposition}
\label{app:greenfun}

\begin{figure}
    \centering
    \includegraphics[width=\columnwidth]{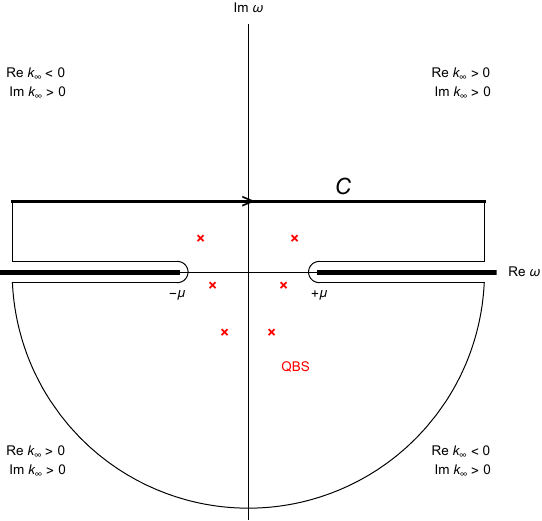}
    \caption{Contour deformation for the retarded Green's function on the QBS Riemann sheet of $k_\infty=\sqrt{\omega^2-\mu^2}$, defined by ${\rm Im}\, k_\infty>0$. The branch cuts are chosen along $(-\infty,-\mu]$ and $[\mu,\infty)$, and the retarded contour $C$ is closed in the lower half $\omega$ plane for $t>t'$. The red crosses denote QBS poles on the sheet, while the branch-cut integrals give the continuum contribution.}
    \label{fig:contour}
\end{figure}

In this appendix, we discuss the retarded Green's function and its mode decomposition. The retarded Green's function satisfies
\be
    \left( \Box_x - \mu^2 \right) G_{a,\rm ret}(x,x') = \frac{\delta^{(4)}(x-x')}{\sqrt{-g(x)}}\,.
\ee
We expand it in Fourier modes and spheroidal harmonics as follows. 
\be
    G_{a,\rm ret} = \sum_{\ell m} \int_C \frac{\dif \omega}{2\pi} e^{-i \omega (t-t')} \tilde{G}_\omega(r,r') e^{i m (\varphi-\varphi')} S_{\ell m}(\theta) S_{\ell m}(\theta')
\ee
The contour $C$ runs from $-\infty+ic$ to $+\infty+ic$ in the complex $\omega$-plane and lies above all poles. The radial Green's function satisfies
\be
    \left( \frac{\dif}{\dif r} \Delta \frac{\dif}{\dif r} + \frac{\left( \omega(r^2+a^2)-a m \right)^2}{\Delta} - \mu^2 r^2 - \lambda_{0 \ell m} \right) \tilde{G}_\omega(r,r') = \delta(r-r'),
\ee
and can be constructed from homogeneous solutions satisfying an ingoing-wave boundary condition at the horizon and an outgoing-wave boundary condition at infinity,
\be
    \tilde{G}_\omega(r,r') = \frac{R^{a, \rm up}(r_>) R^{a, \rm in}(r_<)}{2 i k_\infty A^{a,\rm in}}
\ee
where $r_>=\max(r,r')$ and $r_<=\min(r,r')$. For $t>t'$, the contour may be closed in the lower half-plane, yielding contributions from the branch cuts $(-\infty,-\mu] \cup [\mu,\infty)$ and the poles for which \(A^{a,{\rm in}}=0\) (see Fig.~\ref{fig:contour}). Namely,
\be\label{eq:contour-integral}
\ba 
    & \int_C \dif \omega \frac{R^{a, \rm up}_{\omega}(r_>) R^{a, \rm in}_{\omega}(r_<)}{W(\omega)} \\
    = & \sum_n \frac{R^{a, \rm up}_{\omega_n}(r_>) R^{a, \rm in}_{\omega_n}(r_<)}{\partial_{\omega_n}W(\omega_n)}\\
    & + \left(\int_{-\infty}^{-\mu} + \int_{\mu}^{\infty}\right) \frac{R^{a, \rm up}_{\omega+i \epsilon}(r_>) R^{a, \rm in}_{\omega+i \epsilon}(r_<)}{W(\omega+i \epsilon)} - \frac{R^{a, \rm up}_{\omega-i \epsilon}(r_>) R^{a, \rm in}_{\omega-i\epsilon}(r_<)}{W(\omega-i\epsilon)} \dif \omega.
\ea
\ee
Across the branch cut, one has 
\be
\ba
    R^{a, \rm in}_{\omega+i\epsilon}&=R^{a, \rm in}_{\omega-i\epsilon}, \quad
    R^{a, \rm up}_{\omega+i\epsilon}=(R^{a, \rm up}_{\omega-i\epsilon})^*,\\
    R^{a, \rm in}_{\omega} &= A^{a,\rm in}_{\omega} (R^{a, \rm up}_{\omega})^* + A^{a,\rm out}_{\omega} R^{a, \rm up}_{\omega}.
\ea
\ee
and the boundary condition implies
\be
\ba
    W[R^{a, \rm in},R^{a, \rm up}] &= 2 i k_\infty A^{a,\rm in}, \\
    W[R^{a, \rm in},(R^{a, \rm up})^*] &= -2 i k_\infty A^{a,\rm out}.
\ea
\ee
Hence the second term in the right-hand side of Eq.~\eqref{eq:contour-integral} can be simplified as
\be
\ba
    &\frac{R^{a, \rm up}_{\omega+i \epsilon}(r_>) R^{a, \rm in}_{\omega+i \epsilon}(r_<)}{W(\omega+i \epsilon)} - \frac{R^{a, \rm up}_{\omega-i \epsilon}(r_>) R^{a, \rm in}_{\omega-i\epsilon}(r_<)}{W(\omega-i\epsilon)} \\
    = & \frac{R^{a, \rm up}_{\omega+i \epsilon}(r_>) R^{a, \rm in}_{\omega+i \epsilon}(r_<)}{2 i k_\infty A^{a,\rm in}_{\omega+i \epsilon}} + \frac{(R^{a, \rm up}_{\omega+i \epsilon}(r_>))^* R^{a, \rm in}_{\omega+i\epsilon}(r_<)}{2 i k_\infty A^{a,\rm out}_{\omega+i \epsilon}}\\
    = & \frac{R^{a, \rm in}_{\omega+i \epsilon}(r_<)}{2 i k_\infty} 
        \frac{R^{a, \rm up}_{\omega+i \epsilon}(r_>) A^{a,\rm out}_{\omega+i \epsilon} + (R^{a, \rm up}_{\omega+i \epsilon}(r_>))^* A^{a,\rm in}_{\omega+i \epsilon}}{A^{a,\rm in}_{\omega+i \epsilon} A^{a,\rm out}_{\omega+i \epsilon}}\\
    = & \frac{R^{a, \rm in}_{\omega+i \epsilon}(r_<) R^{a, \rm in}_{\omega+i \epsilon}(r_>)}{2 i k_\infty A^{a,\rm in}_{\omega+i \epsilon} A^{a,\rm out}_{\omega+i \epsilon}}.
\ea
\ee
The discrete spectrum contribution is given by
\be
\ba
    \frac{R^{a, \rm up}_{\omega_n}(r_>) R^{a, \rm in}_{\omega_n}(r_<)}{\partial_{\omega_n}W(\omega_n)} &= \frac{R^{a, \rm up}_{\omega_n}(r_>) R^{a, \rm up}_{\omega_n}(r_<)}{2 i k_\infty A^{a,\rm down}(\omega_n) \partial_{\omega_n} A^{a,\rm in}(\omega_n)} \\
    &= R^a_{n \ell m}(r) R^a_{n \ell m}(r'),
\ea
\ee
where $R^a_{n \ell m}(r)$ satisfies the orthogonality condition given in Eq.~\eqref{eq:orthogonality}, and the second equality has been proved in Ref. \cite{Fu:2025ztk}.

Therefore, the retarded Green's function can be decomposed into a discrete spectrum and a continuum spectrum as
\be
\ba
    G_{a,\rm ret}(x,x') = & \Theta(t-t') \sum_{\ell m} \Bigg[ \sum_{n} e^{-i \omega_n (t-t')} e^{i m (\varphi-\varphi')} S^a_{\ell m}(\theta) S^a_{\ell m}(\theta') \\
    & \qquad \qquad \qquad  \times R^a_{n \ell m}(r) R^a_{n \ell m}(r') \\
    & + \int_\mu^\infty \dif\omega e^{-i \omega (t-t')} e^{i m (\varphi-\varphi')} S^a_{\ell m}(\theta) S^a_{\ell m}(\theta') \\ 
    & \qquad \qquad \qquad \times \frac{R^{a, \rm in}_{\omega}(r) R^{a, \rm in}_{\omega}(r')}{2 i k_\infty A^{a,\rm in}_{\omega} A^{a,\rm out}_{\omega}} \Bigg] + {\rm c.c.}.
\ea
\ee

For a source with harmonic time dependence $\mathcal{S}({\bf r})e^{-i \Omega t}$, the axion field is obtained from the retarded Green's function as
\be
\ba
    \Phi = & \int \dif^4 x' \sqrt{-g}\, G_{a,\rm ret}(x,x') \mathcal{S}({\bf r'}) e^{-i \Omega t'} \\
    = & \sum_{\ell m}\Bigg[ \sum_{n} \int_{-\infty}^{t} \dif t' e^{-i \Omega t'} e^{-i \omega_n (t-t')} \int \dif^3 \mathbf{r}' \sqrt{-g}\, \mathcal{S}({\bf r'}) G^a_{n \ell m}(\bf r,r') \\
    & + \int_{\mu}^{\infty} \dif \omega \int_{-\infty}^{t} \dif t' e^{-i \Omega t'} e^{-i \omega (t-t')} \int \dif^3 \mathbf{r}' \sqrt{-g}\, \mathcal{S}({\bf r'}) G^{a}_{\omega \ell m}(\bf r,r') \Bigg].
\ea
\ee
Here we assume that $\Omega < \mu$ and $\Omega \neq \omega_n$ and only focus on the stationary oscillating solution with frequency $\Omega$. The time integral then gives
\be
    \int_{-\infty}^{t} \dif t' e^{-i (\Omega-\omega) t'} = \frac{e^{-i (\Omega-\omega) t}}{-i (\Omega-\omega)}.
\ee
Therefore, the field solution can be expressed as
\be
\ba
    \Phi = & \sum_{\ell m} \Bigg[ \sum_{n} \frac{e^{-i \Omega t}}{-i (\Omega-\omega_n)} \int \dif^3 \mathbf{r}' \sqrt{-g}\, \mathcal{S}({\bf r'}) e^{i m (\varphi-\varphi')} \\
    & \qquad \qquad \qquad \times S^a_{\ell m}(\theta) S^a_{\ell m}(\theta') R^a_{n \ell m}(r) R^a_{n \ell m}(r') \\
    & + \int_{\mu}^{\infty} \dif \omega \frac{e^{-i \Omega t}}{-i (\Omega-\omega)} \int \dif^3 \mathbf{r}' \sqrt{-g}\, \mathcal{S}({\bf r'}) e^{i m (\varphi-\varphi')} \\
    & \qquad \qquad \qquad \times S^a_{\ell m}(\theta) S^a_{\ell m}(\theta') \frac{R^{a, \rm in}_{\omega}(r) R^{a, \rm in}_{\omega}(r')}{2 i k_\infty A^{a,\rm in}_{\omega} A^{a,\rm out}_{\omega}} \Bigg] + {\rm c.c.}.
\ea
\ee
This reproduces the mode decomposition used in Eq.~\eqref{eq:AxionDecomp}.

\section{Estimating non-Hermitian contributions}
\label{app:nonHerm}

In this appendix, we estimate the non-Hermitian part of the mode decomposition
\be
    \Phi = \sum_{n} \frac{\mathcal{N}_n}{\omega_a - \omega_{n}}, \mathcal{N}_n=\int \dif r' \dif\theta' R_{0 n}(r') S_{0}(\theta') F^{\mu\nu} \tilde{F}_{\mu\nu} \Sigma \sin\theta'
\ee
and show that the dominant non-Hermitian contribution comes from the complex eigenfrequencies. The starting point is the homogeneous axion equation
\be
    \mathcal{H}_0(\omega)\Phi=0 .
\ee
Since we assume $\omega \ll \mu \ll 1$, the effect of the horizon can be treated as a small non-Hermitian perturbation. We write
\be
    \mathcal{H}_0(\omega)
    =
    \mathcal{H}_0^{\rm Herm}(\omega)
    +
    \delta\mathcal{H}_0(\omega),
\ee
where $\mathcal{H}_0^{\rm Herm}$ defines the corresponding Hermitian eigenvalue problem and $\delta\mathcal{H}_0$ encodes the absorptive boundary condition at the horizon.

For a QBS mode labeled by $n\ell m$, we only need the first-order correction to the wave function,
\be
    \Phi_{n\ell m}
    =
    \Phi_{n\ell m}^{(0)}
    +
    \delta\Phi_{n\ell m}.
\ee
At fixed frequency, this correction is obtained by expanding in the complete set of unperturbed modes,
\be
    \delta\Phi_{n\ell m}
    =
    \sum_{\lambda \neq n\ell m}
    c_{\lambda,n\ell m}\Phi_{\lambda}^{(0)} ,
\ee
where the coefficients are 
\be
    c_{\lambda,n\ell m}
    =
    -
    \frac{
    \left\langle \Phi_{\lambda}^{(0)},
    \delta\mathcal{H}_0\left(\omega_{n\ell m}^{(0)}\right)
    \Phi_{n\ell m}^{(0)}
    \right\rangle}
    {(\omega_\lambda-\omega_{n \ell m})
    \left\langle \Phi_{\lambda}^{(0)},\Phi_{\lambda}^{(0)}\right\rangle},
    \qquad \lambda \neq n\ell m .
\ee
For an order-of-magnitude estimate, the numerator in $c_{\lambda,n\ell m}$ is of the same order as the non-Hermitian correction to the QBS eigenfrequency, namely ${\rm Im}\,\omega_n \sim \mu(M\mu)^{4\ell+4}$. The denominator is set by the typical bound-state level spacing, $\Delta\omega \sim \mu(M\mu)^2$. Therefore the mixing coefficient scales as
\be
    c_{\lambda,n\ell m}
    \sim
    \frac{\mu(M\mu)^{4\ell+4}}
    {\mu(M\mu)^2}
    \sim
    (M\mu)^{4\ell+2}.
\ee

The product of the axion source $F^{\mu \nu} \tilde{F}_{\mu \nu}$ and the weight function $\Sigma$ typically scales as $r^{2\ell}$. On the other hand, the non-Hermitian part of the wave functions is localized near the horizon, $r \sim M$, while the Hermitian wave functions are supported near the Bohr radius $1/(M \mu^2)$. The relative non-Hermitian contribution to the overlap integral is therefore
\be
    \frac{\mathcal{N}_n^{\rm non-Herm}}{\mathcal{N}_n^{\rm Herm}} \sim \frac{(M \mu)^{4\ell + 2} \times M^{2\ell}}{(M \mu^2)^{-2\ell}} = (M \mu)^{6\ell +2}.
\ee

By comparison, the relative non-Hermitian contribution from the eigenfrequencies is
\be
    \frac{{\rm Im} \omega_n}{\Re (\omega_a-\omega_n)} \sim \frac{(M \mu)^{4\ell+4} \mu}{\mu} = (M \mu)^{4\ell+4} .
\ee
In our case, $\ell_a \geq 2$. Hence the non-Hermitian contribution from the eigenfrequencies dominates the axion field $\Phi$ at $\mathcal{O}(g_{a\gamma})$.

\bibliography{references}

@book{Chandrasekhar:1985kt,
    author = "Chandrasekhar, Subrahmanyan",
    title = "{The mathematical theory of black holes}",
    isbn = "978-0-19-850370-5",
    year = "1985"
}

@article{Starobinskil:1974nkd,
    author = "Starobinskil, Alexei A. and Churilov, S. M.",
    title = "{Amplification of electromagnetic and gravitational waves scattered by a rotating ''black hole''}",
    journal = "Sov. Phys. JETP",
    volume = "65",
    number = "1",
    pages = "1--5",
    year = "1974"
}

@article{Starobinskii:1973vzb,
    author = "Starobinskii, A. A.",
    title = "{Amplification of waves during reflection from a rotating ''black hole''}",
    journal = "Sov. Phys. JETP",
    volume = "37",
    number = "1",
    pages = "28--32",
    year = "1973"
}

@article{Page:1976df,
    author = "Page, Don N.",
    title = "{Particle Emission Rates from a Black Hole: Massless Particles from an Uncharged, Nonrotating Hole}",
    doi = "10.1103/PhysRevD.13.198",
    journal = "Phys. Rev. D",
    volume = "13",
    pages = "198--206",
    year = "1976"
}

@article{Teukolsky:1974yv,
    author = "Teukolsky, S. A. and Press, W. H.",
    title = "{Perturbations of a rotating black hole. III - Interaction of the hole with gravitational and electromagnetic radiation}",
    doi = "10.1086/153180",
    journal = "Astrophys. J.",
    volume = "193",
    pages = "443--461",
    year = "1974"
}

@article{Teukolsky:1973ha,
    author = "Teukolsky, Saul A.",
    title = "{Perturbations of a rotating black hole. 1. Fundamental equations for gravitational electromagnetic and neutrino field perturbations}",
    doi = "10.1086/152444",
    journal = "Astrophys. J.",
    volume = "185",
    pages = "635--647",
    year = "1973"
}

@article{Cannizzaro:2025vpb,
    author = "Cannizzaro, Enrico and Palleschi, Marco and Sberna, Laura and Brito, Richard and Green, Stephen R.",
    title = "{Excitation of scalar quasinormal modes from boson clouds}",
    eprint = "2512.15878",
    archivePrefix = "arXiv",
    primaryClass = "gr-qc",
    doi = "10.1103/fbwb-1hcz",
    journal = "Phys. Rev. D",
    volume = "113",
    number = "8",
    pages = "083039",
    year = "2026"
}

@article{Green:2022htq,
    author = "Green, Stephen R. and Hollands, Stefan and Sberna, Laura and Toomani, Vahid and Zimmerman, Peter",
    title = "{Conserved currents for a Kerr black hole and orthogonality of quasinormal modes}",
    eprint = "2210.15935",
    archivePrefix = "arXiv",
    primaryClass = "gr-qc",
    doi = "10.1103/PhysRevD.107.064030",
    journal = "Phys. Rev. D",
    volume = "107",
    number = "6",
    pages = "064030",
    year = "2023"
}

@article{Cannizzaro:2023jle,
    author = "Cannizzaro, Enrico and Sberna, Laura and Green, Stephen R. and Hollands, Stefan",
    title = "{Relativistic Perturbation Theory for Black-Hole Boson Clouds}",
    eprint = "2309.10021",
    archivePrefix = "arXiv",
    primaryClass = "gr-qc",
    doi = "10.1103/PhysRevLett.132.051401",
    journal = "Phys. Rev. Lett.",
    volume = "132",
    number = "5",
    pages = "051401",
    year = "2024"
}

@misc{Fu:2025ztk,
    author = "Fu, Lingyun and Omiya, Hidetoshi and Tanaka, Takahiro and Tong, Xi and Wang, Yi and Zhu, Hui-Yu",
    title = "{Quantum Treatment of Black Hole Superradiance}",
    eprint = "2512.06790",
    archivePrefix = "arXiv",
    primaryClass = "gr-qc",
    month = "12",
    year = "2025"
}

@article{Yang:2013shb,
    author = "Yang, Huan and Zhang, Fan and Zimmerman, Aaron and Chen, Yanbei",
    title = "{Scalar Green function of the Kerr spacetime}",
    eprint = "1311.3380",
    archivePrefix = "arXiv",
    primaryClass = "gr-qc",
    doi = "10.1103/PhysRevD.89.064014",
    journal = "Phys. Rev. D",
    volume = "89",
    number = "6",
    pages = "064014",
    year = "2014"
}

@ARTICLE{ZelDovich:1971,
       author = {{Zel'Dovich}, Ya. B.},
        title = "{Generation of Waves by a Rotating Body}",
      journal = {Soviet Journal of Experimental and Theoretical Physics Letters},
        year = 1971,
        month = aug,
       volume = {14},
        pages = {180},
       adsurl = {https://ui.adsabs.harvard.edu/abs/1971JETPL..14..180Z}
}

@article{Press:1972zz,
    author = "Press, William H. and Teukolsky, Saul A.",
    title = "{Floating Orbits, Superradiant Scattering and the Black-hole Bomb}",
    doi = "10.1038/238211a0",
    journal = "Nature",
    volume = "238",
    pages = "211--212",
    year = "1972"
}

@article{Zouros:1979iw,
    author = "Zouros, T. J. M. and Eardley, D. M.",
    title = "{INSTABILITIES OF MASSIVE SCALAR PERTURBATIONS OF A ROTATING BLACK HOLE}",
    doi = "10.1016/0003-4916(79)90237-9",
    journal = "Annals Phys.",
    volume = "118",
    pages = "139--155",
    year = "1979"
}

@article{Detweiler:1980uk,
    author = "Detweiler, Steven L.",
    title = "{KLEIN-GORDON EQUATION AND ROTATING BLACK HOLES}",
    doi = "10.1103/PhysRevD.22.2323",
    journal = "Phys. Rev. D",
    volume = "22",
    pages = "2323--2326",
    year = "1980"
}

@article{Brito:2015oca,
    author = "Brito, Richard and Cardoso, Vitor and Pani, Paolo",
    title = "{Superradiance}: {New Frontiers in Black Hole
Physics}",
    eprint = "1501.06570",
    archivePrefix = "arXiv",
    primaryClass = "gr-qc",
    doi = "10.1007/978-3-319-19000-6",
    journal = "Lect. Notes Phys.",
    volume = "906",
    pages = "pp.1--237",
    year = "2015"
}

@article{Arvanitaki:2009fg,
    author = "Arvanitaki, Asimina and Dimopoulos, Savas and Dubovsky, Sergei and Kaloper, Nemanja and March-Russell, John",
    title = "{String axiverse}",
    eprint = "0905.4720",
    archivePrefix = "arXiv",
    primaryClass = "hep-th",
    doi = "10.1103/PhysRevD.81.123530",
    journal = "Phys. Rev. D",
    volume = "81",
    pages = "123530",
    year = "2010"
}

@article{Arvanitaki:2010sy,
    author = "Arvanitaki, Asimina and Dubovsky, Sergei",
    title = "{Exploring the String Axiverse with Precision Black Hole Physics}",
    eprint = "1004.3558",
    archivePrefix = "arXiv",
    primaryClass = "hep-th",
    doi = "10.1103/PhysRevD.83.044026",
    journal = "Phys. Rev. D",
    volume = "83",
    pages = "044026",
    year = "2011"
}

@article{Arvanitaki:2014wva,
    author = "Arvanitaki, Asimina and Baryakhtar, Masha and Huang, Xinlu",
    title = "{Discovering the QCD Axion with Black Holes and Gravitational Waves}",
    eprint = "1411.2263",
    archivePrefix = "arXiv",
    primaryClass = "hep-ph",
    doi = "10.1103/PhysRevD.91.084011",
    journal = "Phys. Rev. D",
    volume = "91",
    number = "8",
    pages = "084011",
    year = "2015"
}

@article{Xie:2025npy,
    author = "Xie, Ning and Huang, Fa Peng",
    title = "{Self-interaction effects on the Kerr black hole superradiance and their observational implications}",
    eprint = "2503.10347",
    archivePrefix = "arXiv",
    primaryClass = "hep-ph",
    doi = "10.1103/xmhn-cpv4",
    journal = "Phys. Rev. D",
    volume = "112",
    number = "5",
    pages = "055028",
    year = "2025"
}

@article{Baryakhtar:2020gao,
    author = "Baryakhtar, Masha and Galanis, Marios and Lasenby, Robert and Simon, Olivier",
    title = "{Black hole superradiance of self-interacting scalar fields}",
    eprint = "2011.11646",
    archivePrefix = "arXiv",
    primaryClass = "hep-ph",
    doi = "10.1103/PhysRevD.103.095019",
    journal = "Phys. Rev. D",
    volume = "103",
    number = "9",
    pages = "095019",
    year = "2021"
}

@article{Leaver:1985ax,
    author = "Leaver, E. W.",
    title = "{An Analytic representation for the quasi normal modes of Kerr black holes}",
    doi = "10.1098/rspa.1985.0119",
    journal = "Proc. Roy. Soc. Lond. A",
    volume = "402",
    pages = "285--298",
    year = "1985"
}

@article{Baumann:2018vus,
    author = "Baumann, Daniel and Chia, Horng Sheng and Porto, Rafael A.",
    title = "{Probing ultralight bosons with binary black boles}",
    eprint = "1804.03208",
    archivePrefix = "arXiv",
    primaryClass = "gr-qc",
    reportNumber = "DESY-18-060, DESY 18-060",
    doi = "10.1103/PhysRevD.99.044001",
    journal = "Phys. Rev. D",
    volume = "99",
    number = "4",
    pages = "044001",
    year = "2019"
}

@article{leaver1986solutions,
  title={Solutions to a generalized spheroidal wave equation: Teukolsky’s equations in general relativity, and the two-center problem in molecular quantum mechanics},
  author={Leaver, Edward W},
  journal={Journal of mathematical physics},
  volume={27},
  number={5},
  pages={1238--1265},
  year={1986},
  publisher={American Institute of Physics}
}

@article{Brito:2014wla,
    author = "Brito, Richard and Cardoso, Vitor and Pani, Paolo",
    title = "{Black holes as particle detectors: evolution of superradiant instabilities}",
    eprint = "1411.0686",
    archivePrefix = "arXiv",
    primaryClass = "gr-qc",
    doi = "10.1088/0264-9381/32/13/134001",
    journal = "Class. Quant. Grav.",
    volume = "32",
    number = "13",
    pages = "134001",
    year = "2015"
}

@article{Yang:2023vwm,
    author = "Yang, Jing and Huang, Fa Peng",
    title = "{Gravitational waves from axions annihilation through quantum field theory}",
    eprint = "2306.12375",
    archivePrefix = "arXiv",
    primaryClass = "hep-ph",
    doi = "10.1103/PhysRevD.108.103002",
    journal = "Phys. Rev. D",
    volume = "108",
    number = "10",
    pages = "103002",
    year = "2023"
}

@article{Chen:2019fsq,
    author = "Chen, Yifan and Shu, Jing and Xue, Xiao and Yuan, Qiang and Zhao, Yue",
    title = "{Probing Axions with Event Horizon Telescope Polarimetric Measurements}",
    eprint = "1905.02213",
    archivePrefix = "arXiv",
    primaryClass = "hep-ph",
    doi = "10.1103/PhysRevLett.124.061102",
    journal = "Phys. Rev. Lett.",
    volume = "124",
    number = "6",
    pages = "061102",
    year = "2020"
}

@article{Chen:2021lvo,
    author = "Chen, Yifan and Liu, Yuxin and Lu, Ru-Sen and Mizuno, Yosuke and Shu, Jing and Xue, Xiao and Yuan, Qiang and Zhao, Yue",
    title = "{Stringent axion constraints with Event Horizon Telescope polarimetric measurements of M87}",
    eprint = "2105.04572",
    archivePrefix = "arXiv",
    primaryClass = "hep-ph",
    doi = "10.1038/s41550-022-01620-3",
    journal = "Nature Astron.",
    volume = "6",
    number = "5",
    pages = "592--598",
    year = "2022"
}

@article{Chen:2022oad,
    author = "Chen, Yifan and Li, Chunlong and Mizuno, Yosuke and Shu, Jing and Xue, Xiao and Yuan, Qiang and Zhao, Yue and Zhou, Zihan",
    title = "{Birefringence tomography for axion cloud}",
    eprint = "2208.05724",
    archivePrefix = "arXiv",
    primaryClass = "hep-ph",
    doi = "10.1088/1475-7516/2022/09/073",
    journal = "JCAP",
    volume = "09",
    pages = "073",
    year = "2022"
}

@article{Chen:2023vkq,
    author = "Chen, Yifan and Xue, Xiao and Cardoso, Vitor",
    title = "{Black holes as fermion factories}",
    eprint = "2308.00741",
    archivePrefix = "arXiv",
    primaryClass = "hep-ph",
    reportNumber = "DESY-23-109",
    doi = "10.1088/1475-7516/2025/02/035",
    journal = "JCAP",
    volume = "02",
    pages = "035",
    year = "2025"
}

@article{Lyu:2025lue,
    author = "Lyu, Zhen-Hong and Cai, Rong-Gen and Guo, Zong-Kuan and He, Jian-Feng and Liu, Jing",
    title = "{Ring formation from black hole superradiance through repeated particle production on bound orbits}",
    eprint = "2507.03490",
    archivePrefix = "arXiv",
    primaryClass = "gr-qc",
    doi = "10.1103/3r41-xyj3",
    journal = "Phys. Rev. D",
    volume = "112",
    number = "10",
    pages = "104066",
    year = "2025"
}

@article{Zhang:2018kib,
    author = "Zhang, Jun and Yang, Huan",
    title = "{Gravitational floating orbits around hairy black holes}",
    eprint = "1808.02905",
    archivePrefix = "arXiv",
    primaryClass = "gr-qc",
    doi = "10.1103/PhysRevD.99.064018",
    journal = "Phys. Rev. D",
    volume = "99",
    number = "6",
    pages = "064018",
    year = "2019"
}

@article{Zhang:2019eid,
    author = "Zhang, Jun and Yang, Huan",
    title = "{Dynamic signatures of black hole binaries with superradiant clouds}",
    eprint = "1907.13582",
    archivePrefix = "arXiv",
    primaryClass = "gr-qc",
    reportNumber = "Imperial/TP/2019/JZ/02",
    doi = "10.1103/PhysRevD.101.043020",
    journal = "Phys. Rev. D",
    volume = "101",
    number = "4",
    pages = "043020",
    year = "2020"
}

@article{Spieksma:2023vwl,
    author = "Spieksma, Thomas F. M. and Cannizzaro, Enrico and Ikeda, Taishi and Cardoso, Vitor and Chen, Yifan",
    title = "{Superradiance: Axionic couplings and plasma effects}",
    eprint = "2306.16447",
    archivePrefix = "arXiv",
    primaryClass = "gr-qc",
    doi = "10.1103/PhysRevD.108.063013",
    journal = "Phys. Rev. D",
    volume = "108",
    number = "6",
    pages = "063013",
    year = "2023"
}

@article{BekensteinSchiffer:1998,
    author = "Bekenstein, Jacob D. and Schiffer, Marcelo",
    title = "{The Many Faces of Superradiance}",
    eprint = "gr-qc/9803033",
    archivePrefix = "arXiv",
    doi = "10.1103/PhysRevD.58.064014",
    journal = "Phys. Rev. D",
    volume = "58",
    pages = "064014",
    year = "1998"
}

@article{Dolan:2007,
    author = "Dolan, Sam R.",
    title = "{Instability of the Massive Klein--Gordon Field on the Kerr Spacetime}",
    eprint = "0705.2880",
    archivePrefix = "arXiv",
    primaryClass = "gr-qc",
    doi = "10.1103/PhysRevD.76.084001",
    journal = "Phys. Rev. D",
    volume = "76",
    pages = "084001",
    year = "2007"
}

@article{KobayashiTomimatsu:2010,
    author = "Kobayashi, Tsutomu and Tomimatsu, Akira",
    title = "{Superradiant Scattering of Electromagnetic Waves Emitted from Disk around Kerr Black Holes}",
    eprint = "1009.2193",
    archivePrefix = "arXiv",
    primaryClass = "gr-qc",
    doi = "10.1103/PhysRevD.82.084026",
    journal = "Phys. Rev. D",
    volume = "82",
    pages = "084026",
    year = "2010"
}

@article{Zhu:2025enp,
    author = "Zhu, Zhi-Qing and Piao, Yun-Song and Zhang, Jun",
    title = "{Black hole superradiance of interacting multifield}",
    eprint = "2509.10892",
    archivePrefix = "arXiv",
    primaryClass = "gr-qc",
    doi = "10.1103/fscx-79nm",
    journal = "Phys. Rev. D",
    volume = "113",
    number = "6",
    pages = "063048",
    year = "2026"
}
\end{document}